\documentclass[aps,prl,twocolumn,superscriptaddress,showpacs,showkeys,portrait,a4]{revtex4-1}

\usepackage{epsf}
\usepackage{graphicx}
\usepackage{amssymb}
\usepackage{color}
\usepackage{float}
\usepackage{epstopdf}
\usepackage{natbib}
\usepackage{amsmath}

\usepackage{xspace}

\newcommand{\BA}{Ba$_2$YIrO$_6$\xspace}
\newcommand{\SRR}{Sr$_2$YIrO$_6$\xspace}
\newcommand{\BASRR}{(Ba,Sr)$_2$YIrO$_6$\xspace}

\newcommand{\IRS}{Ir$^{6+}$\xspace}
\newcommand{\IRF}{Ir$^{5+}$\xspace}
\newcommand{\IRV}{Ir$^{4+}$\xspace}

\newcommand{\sss}{$S=3/2$\xspace}
\newcommand{\s}{$S=1/2$\xspace}

\newcommand{\jeffe}{$j_{\text{eff}}=1/2$\xspace}

\newcommand{\gl}{$g_{\text{left}}$\xspace}
\newcommand{\gm}{$g_{\text{mid}}$\xspace}
\newcommand{\gr}{$g_{\text{right}}$\xspace}

\newcommand{\CW}{$\theta_{\text{CW}}$\xspace}

\newcommand{\hr}[1]{$\mu_0H_\text{res} =$\,{#1}{\,mT\xspace}}
\newcommand{\dH}{$\Delta H$\xspace}

\begin{document}

\title{Unraveling the Nature of Magnetism of the 5$\boldsymbol{d^4}$  Double Perovskite \BA}
\author{S. Fuchs}
\affiliation{Leibniz Institute for Solid State and Materials Research IFW Dresden, D-01171 Dresden, Germany}
\affiliation{Institut f\"ur Festk\"orper- und Materialphysik, Technische Universit\"at Dresden, D-01062 Dresden, Germany}
\author{T. Dey}
\affiliation{Leibniz Institute for Solid State and Materials Research IFW Dresden, D-01171 Dresden, Germany}
\author{G. Aslan-Cansever}
\affiliation{Leibniz Institute for Solid State and Materials Research IFW Dresden, D-01171 Dresden, Germany}
\affiliation{Institut f\"ur Festk\"orper- und Materialphysik, Technische Universit\"at Dresden, 01062 Dresden, Germany}
\author{A. Maljuk}
\affiliation{Leibniz Institute for Solid State and Materials Research IFW Dresden, D-01171 Dresden, Germany}
\author{S. Wurmehl}
\affiliation{Leibniz Institute for Solid State and Materials Research IFW Dresden, D-01171 Dresden, Germany}
\author{B.\ B\"{u}chner}
\affiliation{Leibniz Institute for Solid State and Materials Research IFW Dresden, D-01171 Dresden, Germany} 
\affiliation{Institut f\"ur Festk\"orper- und Materialphysik, Technische Universit\"at Dresden, D-01062 Dresden, Germany}
\author{V. Kataev}
\affiliation{Leibniz Institute for Solid State and Materials Research IFW Dresden, D-01171 Dresden, Germany}


\begin{abstract}
We report electron spin resonance (ESR) spectroscopy results on the double perovskite \BA. On general grounds, this material is expected to be nonmagnetic due to the strong coupling of the spin and orbital momenta of \IRF (5$d^4$) ions. However, controversial experimental reports on either strong antiferromagnetism with static order at low temperatures or just a weakly paramagnetic behavior have triggered a discussion on the breakdown of the generally accepted scenario of the strongly spin-orbit coupled ground states in the 5$d^4$ iridates and the emergence of a novel exotic magnetic state. Our data evidence that the magnetism of the studied material is solely due to a few percent of  
\IRV and \IRS magnetic defects 
while the regular \IRF sites remain nonmagnetic. Remarkably, the defect \IRS species manifest magnetic correlations in the ESR spectra at $T\lesssim 20$\,K suggesting a long-range character of superexchange in the double perovskites as proposed by recent theories.   
\end{abstract}


\maketitle

{\it Introduction.--}~Complex iridium oxides are attracting since  about 10 years unceasingly large interest in the condensed matter community worldwide due to predictions of exotic ground states in these materials, such as a spin-orbit assisted Mott insulating state, quantum spin liquid phases, Weyl semimetallic behavior, and superconductivity (for reviews see, e.g., \cite{Cao13,Witczak-Krempa:2014aa,Rau:2016aa,Winter17,Cao17}). Such a rich behavior is expected in iridates due to comparable energy scales of spin-orbit coupling (SOC), electronic bandwidths, noncubic crystal fields and local Coulomb interactions $U$. 

In the widely studied Ir-based compounds, such as, e.g., layered perovskites Sr$_2$IrO$_4$ \cite{Kim08} and Sr$_3$Ir$_2$O$_7$ \cite{Kim12a}, honeycomb compounds ${\text{Na}}_{2}{\text{IrO}}_{3}$ \cite{Singh10}, $\alpha$-Li$_2$IrO$_3$ \cite{Singh12} and their three-dimensional analogues $\beta$- and $\gamma$-Li$_2$IrO$_3$ \cite{Biffin14,Takayama15,Modic14}, hyperkagome compound ${\mathrm{Na}}_{4}{\mathrm{Ir}}_{3}{\mathrm{O}}_{8}$ \cite{Okamoto07} and several other materials, the carrier of magnetic moments are \IRV(5$d^5$) ions. Owing to the strong SOC, the spin ($S$) and orbital ($L$) momenta are entangled in \IRV giving rise to the magnetic Kramers doublet characterized by the effective spin \jeffe \cite{Jackeli09}. The complex structure of \jeffe states is in the core of theoretical models predicting exotic magnetic behavior of iridates \cite{Chaloupka10,Chaloupka13}. In contrast, in the case of \IRF(5$d^4$) the $S-L$ coupling should yield a singlet ground state with the total angular momentum $J = 0$, whereas the magnetic $J=1$ triplet lies much higher in energy \cite{Khaliullin_2013} rendering \IRF-based iridates nonmagnetic. 
In this respect, \IRF double-perovskite iridates \SRR, \BA, and their solid solutions have received recently a great deal of interest due to controversial reports on the observation of either strongly antiferromagnetic behavior with static magnetic order at a low temperature \cite{Cao_2014,Terzic17} or only a weak paramagnetism \cite{Ranjbar15,Dey_2016,Phelan_2016,Chen17,Corredor17,Hammerath17}. This has triggered in turn a substantial number of theoretical works developing various scenarios of the breakdown of the $j_{\rm eff}$ description in $4d^4$ and $5d^4$ Mott insulators and its possible relevance to the \IRF double-perovskite iridates \cite{Bhowal15,Meetei15,Pajskr16,Nag16,Kim17,Svoboda17}, in particular, with regard to the proposed mechanism of condensation of $J = 1$ excitons \cite{Khaliullin_2013}.   

In most of the experimental works magnetic properties of \BASRR were characterized by bulk static magnetometry and specific heat measurements which enabled one to estimate the average magnetic moment, the average magnetic exchange coupling strength and to detect a possible transition to the magnetically ordered state. However, considering the controversy of experimental results and theoretical predictions, it is of paramount importance to identify the exact origin of magnetic behavior and to consolidate experimental results with existing theories.   

In this Letter, we report the results of such identification by means of multifrequency electron spin resonance (ESR) spectroscopy. A decisive advantage of ESR is the possibility to separate different contributions to the total static magnetization, to study the dynamics and correlations of different spin species, to determine their spin multiplicity,  and to measure their intrinsic spin susceptibility. The sample used in our ESR study was an assembly of small single crystals of \BA characterized structurally and magnetically in Ref.~\cite{Dey_2016}. It shows a weak magnetic response in the static susceptibility corresponding to the average effective moment $\mu_{\rm eff} = 0.44\mu_{\rm B}$/Ir with no signatures of magnetic order down to 0.4\,K. A rich ESR spectrum comprising several lines was observed. A careful analysis of the frequency- and temperature-dependent ESR data yields several important findings: (i) the total concentration of magnetic centers contributing to ESR signals amounts to $\sim 4$\,\% of all Ir ions; (ii) the major part of them can be unambiguously identified with \IRV(5$d^5$) and  \IRS(5$d^3$) magnetic ions. In particular, \IRS spin-only centers with \sss show a typical triplet fine structure in the ESR spectrum and a characteristic shift of the spectroscopic $g$ factor; (iv) \IRS spin centers exhibit correlated behavior below $\sim 20$\,K. This enables a definitive conclusion that the magnetism of \BA\   is not related to the conjectured breakdown of the $J = 0$ state of the regular \IRF(5$d^4$) lattice in this material and the occurrence of a weak magnetic moment on every \IRF(5$d^4$) site  
but is rather due to different kinds of interacting paramagnetic defects which could even order magnetically at a low temperature if their concentration exceeds a certain threshold level.

\begin{figure}[t]
	\centering
	\includegraphics[clip,width=0.8\columnwidth]{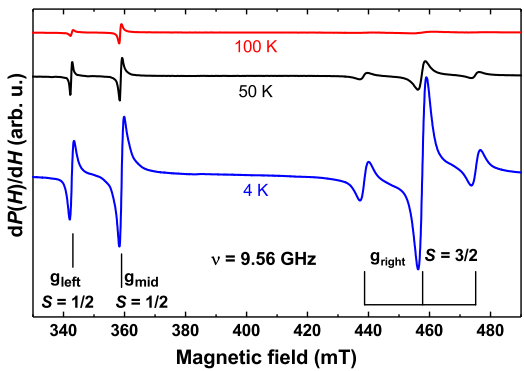}
	\caption{ESR spectra (field derivatives of absorption) at the X-band frequency $\nu = 9.56$\,GHz at three selected temperatures. The lines at $\sim 340$ and $\sim 360$\,mT, and a triplet structure centered around $\sim 460$\,mT are labeled as \gl, \gm, and \gr, respectively, with the spin values assigned to each line. 
		} \label{Spectra10GHz}
\end{figure}

{\it Results.--}~Representative ESR spectra of \BA at different temperatures measured at frequency $\nu = 9.56$\,GHz with a Bruker EMX X-band spectrometer are presented in Fig.~\ref{Spectra10GHz}. Each spectrum comprises a set of sharp resonance lines in the field range $0.33 - 0.49$\,T. The triplet set of lines at the high field side is composed of the main peak at a resonance field \hr{458}, which is accompanied by two satellites at the left and right sides of the main peak. At the low field side, there are two lines at \hr{343} and 359\,mT. Assuming the simple paramagnetic ESR resonance condition $h\nu = g\mu_{\rm B}\mu_0H_{\rm res}$ one obtains the effective $g$ factors for the left, middle, and high field side peaks of \gl\,=\,2.00, \gm\,=\,1.90 and \gr\,=\,1.49, respectively. Here $h$ is the Planck constant, $\mu_{\rm B}$ is the Bohr magneton, and $\mu_0$ is the vacuum permeability.      

\begin{figure}
	\centering
	\includegraphics[clip,width=0.85\columnwidth]{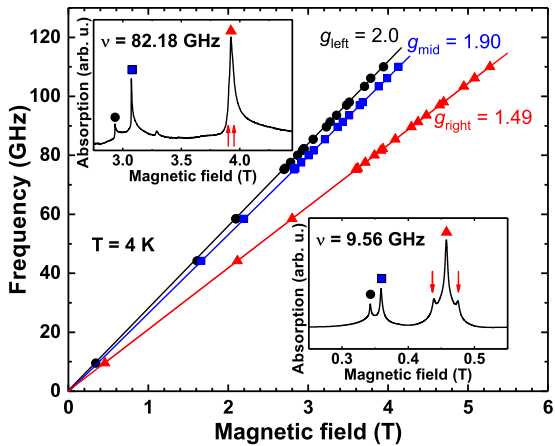}
	\caption{Frequency $\nu$ vs resonance field $H_{\rm res}$ dependence of the peaks in the ESR spectrum (data points). Solid lines are fits to the relation $h\nu = g\mu_{\rm B}\mu_0H_{\rm res}$ yielding the $g$ factor values as indicated in the plot. The insets show spectra at two selected frequencies. The spectrum at 9.56\,GHz was obtained by integration of the absorption derivative spectrum (cf. Fig.~\ref{Spectra10GHz}). Arrows in the upper inset indicate the expected positions of the satellites of the \gr peak in the spectrum at 82.18\,GHz which are resolved at 9.5\,GHz (lower inset).} \label{FreqDepend}
\end{figure}

The single lines \gl and \gm in the ESR spectrum can be straightforwardly assigned to magnetic species carrying the spin $S=1/2$. To identify the spin centers giving rise to the triplet structure around 460\,mT ESR measurements at higher excitation frequencies have been performed with a homemade spectrometer \cite{Golze_2006} equipped with the PNA-X network analyzer from Keysight Technology and a 16\,T superconducting magnet system from Oxford Instruments. In Fig.~\ref{FreqDepend}, the $\nu - H_{\rm res}$ diagram of the resonance modes is shown.  The resonance branches $\nu(H_{\rm res})$ are linear in field. Their slopes $\partial\nu/\partial H$ yield the $g$ factors that nicely agree with the result obtained at $\nu = 9.56$\,GHz. As the $g$ values for the three main lines are different, the spacing between the lines in the spectrum progressively increases with increasing $\nu$. Remarkably, this is not the case for the satellites of the \gr peak. Being resolved at $\sim 10$\,GHz, at higher frequencies they remain hidden under the broadened main peak, suggesting that this group of lines is characterized by the same $g$ factor \gr\,=\,1.49. Such a triplet structure typically arises from magnetic species carrying spin $S = 3/2$. In a solid, the $(2S+1)$-fold degeneracy of the spin levels can be partially lifted in a zero magnetic field due to a combined action of the crystal field (CF) and the spin-orbit coupling. The splitting of these levels giving rise to a fine structure of the ESR signal  can be described by the Hamiltonian \cite{abragam}
\begin{eqnarray}
\label{Hamiltion}
\mathbf{\mathcal{H}}= \mu_{\rm B} \vec{S}\cdot \mathbf{g} \cdot \vec{H}+\vec{S}\cdot \mathbf{D} \cdot \vec{S}.
\end{eqnarray}
Here, the fist and second terms account for the Zeeman interaction with the magnetic field and the interaction with the crystal field, respectively. In a simple case of uniaxial symmetry, the CF tensor {\bf D} reduces to a scalar, and the second term of (\ref{Hamiltion}) simplifies to    
\begin{eqnarray}
\label{HamiltionEasy}
\vec{S}\cdot \mathbf{D} \cdot \vec{S}&=&D\left[S_{z}^{2}-S(S+1)/3\right]. 
\end{eqnarray}
It follows from (\ref{HamiltionEasy}) that the Kramers doublets $|\pm 1/2\rangle$ and $|\pm 3/2\rangle$ of the $S = 3/2$ spin multiplet are separated in energy by $2D$. This gives rise to a "fine-structure" of the ESR spectrum consisting of the main peak due to the resonance transition $|+ 1/2\rangle \leftrightarrow |- 1/2\rangle$ and two weaker in intensity satellites $|\pm 1/2\rangle \leftrightarrow |\pm 3/2\rangle$ with a frequency-independent offset $\pm D$ from the central line.  

\begin{figure}
	\centering
	\includegraphics[clip,width=0.85\columnwidth]{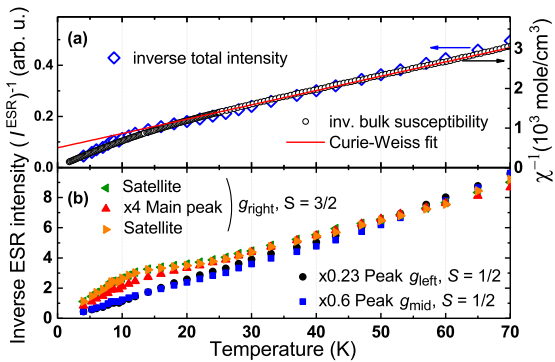}
	\caption{$T$ dependence of the inverse ESR intensity $1/I^{\rm ESR}$ at $\nu = 9.56$\,GHz and its comparison with the static bulk susceptibility $\chi$: (a) total ESR intensity (diamonds, left scale), bulk susceptibility (circles, right scale), and its Curie-Weiss fit $\chi^{-1} = [\chi_0 + C/(T - \theta_{\rm CW})]^{-1}$ with the antiferromagnetic Curie-Weiss temperature $\theta_{\rm CW} = - 16$\,K and the Curie constant $C =0.0294$\,cm$^3$K/mol corresponding to the effective magnetic moment $\mu_{\rm eff} = 0.48\mu_{\rm B}$/Ir (solid line, right scale); (b) intensities of individual lines $1/I^{\rm ESR}_{\rm i}$. For better comparison, the data are scaled as indicated in the legend.    
		} \label{Suscept}
\end{figure}

Since the integrated intensity of an ESR signal $I^{\rm ESR}$ is proportional to the static susceptibility $\chi$ of the resonating spins \cite{abragam}, it can be compared with the bulk susceptibility measurements (Fig.~\ref{Suscept}). The $T$ dependence of the total intensity $I^{\rm ESR}_{\rm tot}$ of all lines in the ESR spectrum of \BA  agrees very well with the static magnetic data $\chi(T)$ [Fig.~\ref{Suscept}(a)], suggesting that the same spins are probed by ESR and static magnetic measurements. In particular, $I^{\rm ESR}_{\rm tot}$ follows the Curie-Weiss law at higher temperatures and, similar to $\chi(T)$, deviates from it below $T\sim 15 -20$\,K signaling the onset of the correlated regime for the resonating spins.     
The signal \gr makes the major contribution to $I^{\rm ESR}_{\rm tot}$ of $\sim$\,73\,\%, whereas signals \gl and \gm contribute to a much lesser extent (see Table~\ref{ESRparam}). To estimate the absolute concentration $n_{\rm i}$ of the spins contributing to the respective signals their intensities were calibrated against a reference sample, a single crystal of Al$_2$O$_3$ doped with a well-defined, small concentration of Cr$^{3+}$ ions (for details, see Refs.~\cite{Rubin1,Rubin2}). The analysis (summarized in Table~\ref{ESRparam}) reveals the total concentration of spins $\sum_{\rm i}n_{\rm i}$ contributing to the ESR spectrum of about 4\,\% per unit cell of \BA. This value is similar to the spin concentration evaluated from the analysis of the static magnetic data \cite{Dey_2016}.    
\begin{table}[htbp] 
	\begin{center}	
		\caption{Parameters of the lines in the ESR spectrum of \BA: g factor, spin value $S$, Curie-Weiss temperature obtained from ESR intensities of individual lines \CW, relative spectral weights  of the signals $I^{\rm ESR}_{\rm i}$, absolute concentration of spins per unit cell $n_{\rm i}$, and the orbital reduction factor $k$.}
		\resizebox{\columnwidth}{!}{
	
			\begin{tabular}{|c|c|c|c|c|c|c|  }\hline   %
				Signal & $g$ factor & $S$ & \CW\,(K) &$I^{\rm ESR}_{\rm i}/I^{\rm ESR}_{\rm tot}$\,(\%)& $n_{\rm i}$ &  $k$ \\ \hline \hline
				\gl & 2.00 & 1/2 & $\sim$\,-2 & $\sim$\,7 & $\sim$\,0.6	&1	\\ \hline 
				\gm  & 1.90 & 1/2 &$\sim$\,-2 & $\sim$\,20 & $\sim$\,1.7& 	0.93	\\ \hline 
				\gr & 1.49 & 3/2 & $\sim$\,-10 & $\sim$\,73 & $\sim$\,1.9 &	0.4	\\ \hline
		\end{tabular}
	}
			\label{ESRparam}
	\end{center}
\end{table}

As can be concluded from the comparison of Figs.~\ref{Suscept}(a) and (b) the $S = 3/2$ centers which give rise to the ESR signal \gr are mainly responsible for the deviation of the spin susceptibility from the paramagnetic Curie-Weiss dependence at low temperatures, have the largest Curie-Weiss temperature \CW (Table~\ref{ESRparam}), and, thus, are "more correlated" than other spin species contributing to the signals \gl and \gm. Additional evidence for magnetic correlations at low $T$ comes from the temperature dependence of the ESR linewidth \dH. Concomitantly with the deviation of $\chi(T)$ from the Curie-Weiss law the linewidth begins to grow below $\sim 20$\,K, indicating the onset of the critical regime characterized by the slowing down of the timescale of spin-spin correlations and a growth of their spatial extension \cite{Huber72}. At higher $T$, \dH becomes constant for \gl and \gm lines, which is typical for \s systems with the dominant Heisenberg isotropic exchange interaction in the noncritical regime \cite{Huber72}. Interestingly, for the \gr line, \dH starts to increase above $\sim 35$\,K again, which is indeed characteristic for \sss systems where the phonon modulation of the crystal field potential gives rise to a $T$-dependent spin-lattice relaxation at elevated temperatures \cite{Huber75}.    

\begin{figure}
	\centering
	\includegraphics[clip,width=0.80\columnwidth]{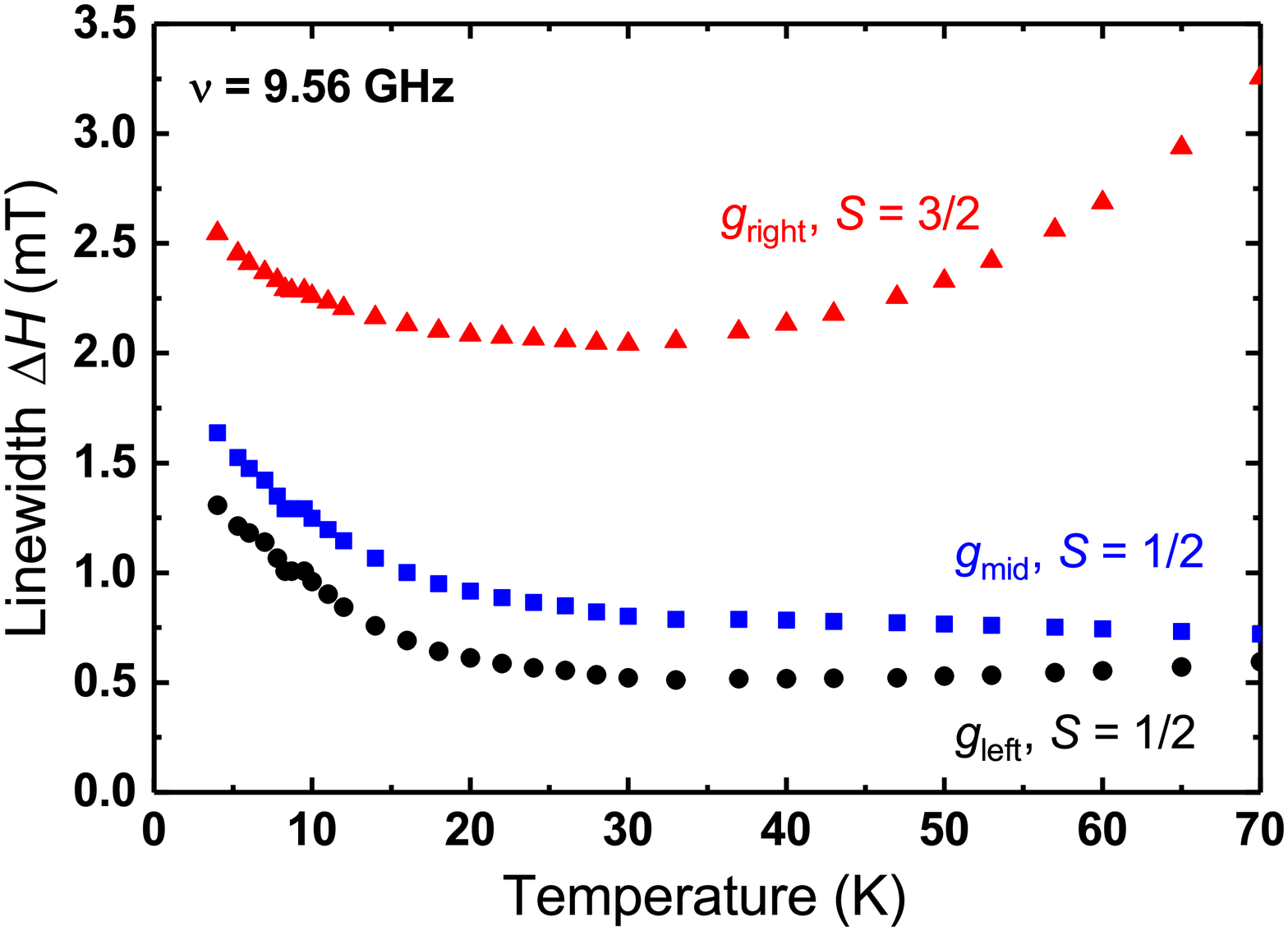}
	\caption{$T$ dependence of the width \dH of the ESR signals \gl, \gm and \gr (main peak) at $\nu = 9.56$\,GHz.       
	} \label{Linewidth}
\end{figure}

{\it Discussion.--}~The small number of magnetic centers contributing to the static magnetization and to ESR spectra of the studied samples of \BA enable a conclusion that the majority of \IRF(5$d^4$) ions in this compound is in the expected nonmagnetic $J = 0$ state. Thus, the observed magnetic response  can be due to the defect Ir sites in the structure with possibly different valences which are likely to occur in a real material. In this respect particular striking is the observation of the \sss centers. Among common oxidation states of Ir, only \IRS(5$d^3$) has such a spin value. Three 5$d$ electrons evenly occupy three orbitals of the $t_{\rm 2g}$ set rendering \IRS a spin-only \sss ion with no orbital momentum in first order. A classical example of the fine-structure triplet ESR spectrum of an \sss paramagnetic center is the ESR response of Cr$^{3+}$(3$d^3$) ions in a octahedral ligand coordination \cite{abragam}. It is characterized by a $g$ factor very close to the spin-only value $g_{\rm s} = 2$ due to the absence of the orbital contribution. A small negative shift $\sim - 0.05$ from $g_{\rm s}$ due to the second-order spin-orbit coupling effect is parametrized in the perturbation theory as \cite{abragam}       
\begin{eqnarray}
g_{||} \approx g_{\perp} =2- 8 k \lambda/\Delta.
\label{gspin32}
\end{eqnarray}
Here, indexes $||$ and $\perp$ denote parallel and perpendicular orientation, respectively, of the applied magnetic field with respect to the symmetry axis of the octahedron, $\lambda$ is the SOC constant, $\Delta$ is the energy difference between the $t_{\rm 2g}$ and $e_{\rm g}$ sets of orbitals, and $k\leq 1$ is the so-called orbital reduction factor which accounts for the covalent character of the metal-ligand bonds ($k = 1$ for ionic bond). A substantially larger negative $g$ shift of $- 0.51$ of the \gr signal can be consistently explained by a combined effect of a much stronger spin-orbit coupling in 5$d$ Ir as compared to a 3$d$ ion and the counteracting effect of the strongly covalent character of  Ir - O  bonds of the highly oxidized \IRS \cite{Choy_1995}. Indeed, with $\lambda \approx 0.5$\,eV \cite{Clancy12}, $\Delta \approx 3.2$\,eV \cite{Choy_1995}, and  \gr\,=\,1.49, one obtains from (\ref{gspin32}) a rather small value of $k = 0.4$ as is generally expected for 5$d$ elements in a high oxidation state (see, e.g., \cite{Allen72,Magnuson84,Reynolds92}). 

The \s  ESR line \gm is characterized by a smaller but still a significant negative shift of the $g$ factor from $g_{\rm s} = 2$. This signal can be assigned to \IRV (5$d^5$) centers with the effective spin \jeffe  covalently bonded with the ligands. Since the spin and orbital momenta are entangled in the \IRV iridates \cite{Jackeli09}, the $g$ factor is generally anisotropic if the ligand coordination deviates from an ideal octahedral symmetry \cite{abragam,Bogdanov15}:
%
%
%
\begin{flalign}
&g_{||} = (g_s+2k)\cos^2\alpha-g_s \sin^2 \alpha;    \\
&g_{\perp} = \sqrt{2}k \sin 2\alpha +g_s \sin^2 \alpha;\mbox{\ } \tan 2 \alpha = (2\sqrt{2}\lambda)/(\lambda-2\delta). \nonumber
\label{gspin12}
\end{flalign}
Here, $\delta$ is the energy difference between ($xz$, $yz$) and $xy$ orbitals of the $t_{\rm 2g}$ set arising due to uniaxial distortion. For small distortions $\delta \ll \lambda$, $g_{||} \approx g_{\perp} \approx (g_{\rm s}+ 4k)/3$. With \gm\,=\,1.9, one obtains $k = 0.93$. The larger value of $k$ as compared to the ESR line \gr is fully consistent with the expected smaller covalency of the \IRV - O bond due to a lower oxidation state of the metal ion \cite{Choy_1995}. Similar results were reported for \IRV centers in other hosts with nearly cubic local symmetry \cite{Griffiths54,Schirmer84}. Finally, the smallest in intensity ESR signal \gl = $g_{\rm s} =2.0$ presumably arises from some \s defect centers without sizable covalency effects ($k = 1$). 
{Given a very small concentration of $\sim 0.6$\,\% of these spin species (Table~\ref{ESRparam}), they could be tentatively assigned to stable radical centers localized at structural imperfections often found in oxide materials (see, e.g., \cite{Ganduglia07}).
Both \gl and \gm centers 
carrying a small spin \s can be considered as the spin probes sensitive to magnetic correlations in the subsystem of the interacting \IRS \sss sites in \BA. This explains pronounced low-$T$ upturns of the linewidths of the \gl and \gm signals (Fig.~\ref{Linewidth}) most likely arising due to inhomogeneous quasistatic local fields developing in the \sss correlated network below $\sim 15 - 20$\,K. 
Here one can trace an analogy with an inhomogeneous broadening of a nuclear magnetic resonance signal of a magnetic solid due to the enhancement of electron spin correlations (see, e.g., \cite{Iakovleva15}).
                  
The exact reasons for the occurrence of \IRS \sss centers that appear to be mainly responsible for the unexpected  correlated magnetism of  \BA have yet to be elucidated. Since the partial concentrations $n_{\rm i}$ of \gm and \gr centers are close (Table~\ref{ESRparam}), one thinkable scenario could be a partial static charge disproportionation \IRF\,$\Rightarrow$\,\IRV\,+\,\IRS. Indeed, since the ESR intensity is proportional to the square of the effective moments of the spins contributing to a given resonance line $I^{\rm ESR}\sim \mu_{\rm eff}^{2} = g^{2}S(S+1)\mu_{\rm B}^{2}$ \cite{abragam}, then with $g$ factors from Table~\ref{ESRparam} one obtains the ratio $I^{\rm ESR}({\rm Ir}^{6+}, S = 3/2)/I^{\rm ESR}({\rm Ir}^{4+}, j_{\rm eff} = 1/2)$\,=\,3.1, which is close to the intensity ratio of the \gm and the \gr signals of 3.65 (Table~\ref{ESRparam}). Additionally, \IRS sites could probably arise due to oxygen excess and/or Ba deficiency. The fact that, despite a relatively small concentration of Ir-related defects, they exhibit spin-correlated behavior below $\sim 20$\,K implies significance of long superexchange paths involving several oxygen bridges. This supports theoretical scenarios of the long-range character of magnetic interactions in the 5$d$ double perovskites \cite{Ou14,Kanungo16} with the active role of nonmagnetic cations, such as Y$^{3+}$, as mediators of exchange \cite{Kanungo16}. Furthermore, if to consider the antisite Y\,$\leftrightarrow$\,Ir disorder found in \BA \cite{Chen17}, the Ir-related defect spin centers might occur also at the Y site. In this situation, as our numerical simulations show \cite{Suppl}, magnetic defects even in a moderate concentration of $\sim 5 - 8$\,\% could form extended correlated clusters. 

{\it Conclusions.--} Our multifrequency ESR experiments on the pentavalent iridium double perovskite \BA reveal different paramagnetic centers with the total concentration of $\sim 4$\,\% and completely explain the overall static magnetic response. The major contribution can be unambiguously assigned to the defect \IRS \sss sites which show clear signatures of magnetic interaction at temperatures below $\sim 20$\,K. These experimental results give evidence that the regular \IRF (5$d^{4}$) ions remain in the nonmagnetic $J = 0$ state in \BA, which questions, in general, the scenario of the breakdown of the spin-orbit coupled $j_{\rm eff}$ states in the 5$d^{4}$ double perovskite iridates and the occurrence of a weak magnetic moment on every \IRF(5$d^4$) site. In turn, our findings highlight the relevance of the long-range magnetic interactions in 5$d$ double perovskites proposed in recent theoretical models which might be even responsible for the magnetic order of defect Ir-based spin centers in \BA if their concentration exceeds a certain threshold value.                   

This work was supported by the Deutsche Forschungsgemeinschaft (DFG) through Grant No. KA 1694/8-1, SFB 1143 (Project B01), and Emmy Noether Grant No. WU 595/3-3.

\bibliography{Ba2YIrO6BIBLO}

\begin{thebibliography}{50}%
\makeatletter
\providecommand \@ifxundefined [1]{%
 \@ifx{#1\undefined}
}%
\providecommand \@ifnum [1]{%
 \ifnum #1\expandafter \@firstoftwo
 \else \expandafter \@secondoftwo
 \fi
}%
\providecommand \@ifx [1]{%
 \ifx #1\expandafter \@firstoftwo
 \else \expandafter \@secondoftwo
 \fi
}%
\providecommand \natexlab [1]{#1}%
\providecommand \enquote  [1]{``#1''}%
\providecommand \bibnamefont  [1]{#1}%
\providecommand \bibfnamefont [1]{#1}%
\providecommand \citenamefont [1]{#1}%
\providecommand \href@noop [0]{\@secondoftwo}%
\providecommand \href [0]{\begingroup \@sanitize@url \@href}%
\providecommand \@href[1]{\@@startlink{#1}\@@href}%
\providecommand \@@href[1]{\endgroup#1\@@endlink}%
\providecommand \@sanitize@url [0]{\catcode `\\12\catcode `\$12\catcode
  `\&12\catcode `\#12\catcode `\^12\catcode `\_12\catcode `\%12\relax}%
\providecommand \@@startlink[1]{}%
\providecommand \@@endlink[0]{}%
\providecommand \url  [0]{\begingroup\@sanitize@url \@url }%
\providecommand \@url [1]{\endgroup\@href {#1}{\urlprefix }}%
\providecommand \urlprefix  [0]{URL }%
\providecommand \Eprint [0]{\href }%
\providecommand \doibase [0]{http://dx.doi.org/}%
\providecommand \selectlanguage [0]{\@gobble}%
\providecommand \bibinfo  [0]{\@secondoftwo}%
\providecommand \bibfield  [0]{\@secondoftwo}%
\providecommand \translation [1]{[#1]}%
\providecommand \BibitemOpen [0]{}%
\providecommand \bibitemStop [0]{}%
\providecommand \bibitemNoStop [0]{.\EOS\space}%
\providecommand \EOS [0]{\spacefactor3000\relax}%
\providecommand \BibitemShut  [1]{\csname bibitem#1\endcsname}%
\let\auto@bib@innerbib\@empty
\bibitem [{\citenamefont {Cao}\ and\ \citenamefont {DeLong}(2013)}]{Cao13}%
  \BibitemOpen
  \bibfield  {author} {\bibinfo {author} {\bibfnamefont {G.}~\bibnamefont
  {Cao}}\ and\ \bibinfo {author} {\bibfnamefont {L.~E.}\ \bibnamefont
  {DeLong}},\ }\href@noop {} {\emph {\bibinfo {title} {Frontiers of 4d- and
  5d-Transition Metal Oxides}}}\ (\bibinfo  {publisher} {World Scientific},\
  \bibinfo {address} {Singapore},\ \bibinfo {year} {2013})\BibitemShut
  {NoStop}%
\bibitem [{\citenamefont {Witczak-Krempa}\ \emph {et~al.}(2014)\citenamefont
  {Witczak-Krempa}, \citenamefont {Chen}, \citenamefont {Kim},\ and\
  \citenamefont {Balents}}]{Witczak-Krempa:2014aa}%
  \BibitemOpen
  \bibfield  {author} {\bibinfo {author} {\bibfnamefont {W.}~\bibnamefont
  {Witczak-Krempa}}, \bibinfo {author} {\bibfnamefont {G.}~\bibnamefont
  {Chen}}, \bibinfo {author} {\bibfnamefont {Y.~B.}\ \bibnamefont {Kim}}, \
  and\ \bibinfo {author} {\bibfnamefont {L.}~\bibnamefont {Balents}},\ }\href
  {\doibase 10.1146/annurev-conmatphys-020911-125138} {\bibfield  {journal}
  {\bibinfo  {journal} {Annu. Rev. Condens. Matter Phys.}\ }\textbf {\bibinfo
  {volume} {5}},\ \bibinfo {pages} {57} (\bibinfo {year} {2014})}\BibitemShut
  {NoStop}%
\bibitem [{\citenamefont {Rau}\ \emph {et~al.}(2016)\citenamefont {Rau},
  \citenamefont {Lee},\ and\ \citenamefont {Kee}}]{Rau:2016aa}%
  \BibitemOpen
  \bibfield  {author} {\bibinfo {author} {\bibfnamefont {J.~G.}\ \bibnamefont
  {Rau}}, \bibinfo {author} {\bibfnamefont {E.~K.-H.}\ \bibnamefont {Lee}}, \
  and\ \bibinfo {author} {\bibfnamefont {H.-Y.}\ \bibnamefont {Kee}},\ }\href
  {\doibase 10.1146/annurev-conmatphys-031115-011319} {\bibfield  {journal}
  {\bibinfo  {journal} {Annu. Rev. Condens. Matter Phys.}\ }\textbf {\bibinfo
  {volume} {7}},\ \bibinfo {pages} {195} (\bibinfo {year} {2016})}\BibitemShut
  {NoStop}%
\bibitem [{\citenamefont {Winter}\ \emph {et~al.}(2017)\citenamefont {Winter},
  \citenamefont {Tsirlin}, \citenamefont {Daghofer}, \citenamefont {van~den
  Brink}, \citenamefont {Singh}, \citenamefont {Gegenwart},\ and\ \citenamefont
  {Valent\'{\i}}}]{Winter17}%
  \BibitemOpen
  \bibfield  {author} {\bibinfo {author} {\bibfnamefont {S.~M.}\ \bibnamefont
  {Winter}}, \bibinfo {author} {\bibfnamefont {A.~A.}\ \bibnamefont {Tsirlin}},
  \bibinfo {author} {\bibfnamefont {M.}~\bibnamefont {Daghofer}}, \bibinfo
  {author} {\bibfnamefont {J.}~\bibnamefont {van~den Brink}}, \bibinfo {author}
  {\bibfnamefont {Y.}~\bibnamefont {Singh}}, \bibinfo {author} {\bibfnamefont
  {P.}~\bibnamefont {Gegenwart}}, \ and\ \bibinfo {author} {\bibfnamefont
  {R.}~\bibnamefont {Valent\'{\i}}},\ }\href
  {http://stacks.iop.org/0953-8984/29/i=49/a=493002} {\bibfield  {journal}
  {\bibinfo  {journal} {J. Phys: Condense Matter}\ }\textbf {\bibinfo {volume}
  {29}},\ \bibinfo {pages} {493002} (\bibinfo {year} {2017})}\BibitemShut
  {NoStop}%
\bibitem [{\citenamefont {Cao}\ and\ \citenamefont
  {Schlottmann}(2018)}]{Cao17}%
  \BibitemOpen
  \bibfield  {author} {\bibinfo {author} {\bibfnamefont {G.}~\bibnamefont
  {Cao}}\ and\ \bibinfo {author} {\bibfnamefont {P.}~\bibnamefont
  {Schlottmann}},\ }\href {http://stacks.iop.org/0034-4885/81/i=4/a=042502}
  {\bibfield  {journal} {\bibinfo  {journal} {Rep. Prog. Phys.}\ }\textbf
  {\bibinfo {volume} {81}},\ \bibinfo {pages} {042502} (\bibinfo {year}
  {2018})}\BibitemShut {NoStop}%
\bibitem [{\citenamefont {Kim}\ \emph {et~al.}(2008)\citenamefont {Kim},
  \citenamefont {Jin}, \citenamefont {Moon}, \citenamefont {Kim}, \citenamefont
  {Park}, \citenamefont {Leem}, \citenamefont {Yu}, \citenamefont {Noh},
  \citenamefont {Kim}, \citenamefont {Oh}, \citenamefont {Park}, \citenamefont
  {Durairaj}, \citenamefont {Cao},\ and\ \citenamefont {Rotenberg}}]{Kim08}%
  \BibitemOpen
  \bibfield  {author} {\bibinfo {author} {\bibfnamefont {B.~J.}\ \bibnamefont
  {Kim}}, \bibinfo {author} {\bibfnamefont {H.}~\bibnamefont {Jin}}, \bibinfo
  {author} {\bibfnamefont {S.~J.}\ \bibnamefont {Moon}}, \bibinfo {author}
  {\bibfnamefont {J.-Y.}\ \bibnamefont {Kim}}, \bibinfo {author} {\bibfnamefont
  {B.-G.}\ \bibnamefont {Park}}, \bibinfo {author} {\bibfnamefont {C.~S.}\
  \bibnamefont {Leem}}, \bibinfo {author} {\bibfnamefont {J.}~\bibnamefont
  {Yu}}, \bibinfo {author} {\bibfnamefont {T.~W.}\ \bibnamefont {Noh}},
  \bibinfo {author} {\bibfnamefont {C.}~\bibnamefont {Kim}}, \bibinfo {author}
  {\bibfnamefont {S.-J.}\ \bibnamefont {Oh}}, \bibinfo {author} {\bibfnamefont
  {J.-H.}\ \bibnamefont {Park}}, \bibinfo {author} {\bibfnamefont
  {V.}~\bibnamefont {Durairaj}}, \bibinfo {author} {\bibfnamefont
  {G.}~\bibnamefont {Cao}}, \ and\ \bibinfo {author} {\bibfnamefont
  {E.}~\bibnamefont {Rotenberg}},\ }\href@noop {} {\bibfield  {journal}
  {\bibinfo  {journal} {Phys. Rev. Lett.}\ }\textbf {\bibinfo {volume} {101}},\
  \bibinfo {pages} {076402} (\bibinfo {year} {2008})}\BibitemShut {NoStop}%
\bibitem [{\citenamefont {Kim}\ \emph {et~al.}(2012)\citenamefont {Kim},
  \citenamefont {Choi}, \citenamefont {Kim}, \citenamefont {Mitchell},
  \citenamefont {Jackeli}, \citenamefont {Daghofer}, \citenamefont {van~den
  Brink}, \citenamefont {Khaliullin},\ and\ \citenamefont {Kim}}]{Kim12a}%
  \BibitemOpen
  \bibfield  {author} {\bibinfo {author} {\bibfnamefont {J.~W.}\ \bibnamefont
  {Kim}}, \bibinfo {author} {\bibfnamefont {Y.}~\bibnamefont {Choi}}, \bibinfo
  {author} {\bibfnamefont {J.}~\bibnamefont {Kim}}, \bibinfo {author}
  {\bibfnamefont {J.~F.}\ \bibnamefont {Mitchell}}, \bibinfo {author}
  {\bibfnamefont {G.}~\bibnamefont {Jackeli}}, \bibinfo {author} {\bibfnamefont
  {M.}~\bibnamefont {Daghofer}}, \bibinfo {author} {\bibfnamefont
  {J.}~\bibnamefont {van~den Brink}}, \bibinfo {author} {\bibfnamefont
  {G.}~\bibnamefont {Khaliullin}}, \ and\ \bibinfo {author} {\bibfnamefont
  {B.~J.}\ \bibnamefont {Kim}},\ }\href {\doibase
  10.1103/PhysRevLett.109.037204} {\bibfield  {journal} {\bibinfo  {journal}
  {Phys. Rev. Lett.}\ }\textbf {\bibinfo {volume} {109}},\ \bibinfo {pages}
  {037204} (\bibinfo {year} {2012})}\BibitemShut {NoStop}%
\bibitem [{\citenamefont {Singh}\ and\ \citenamefont
  {Gegenwart}(2010)}]{Singh10}%
  \BibitemOpen
  \bibfield  {author} {\bibinfo {author} {\bibfnamefont {Y.}~\bibnamefont
  {Singh}}\ and\ \bibinfo {author} {\bibfnamefont {P.}~\bibnamefont
  {Gegenwart}},\ }\href {\doibase 10.1103/PhysRevB.82.064412} {\bibfield
  {journal} {\bibinfo  {journal} {Phys. Rev. B}\ }\textbf {\bibinfo {volume}
  {82}},\ \bibinfo {pages} {064412} (\bibinfo {year} {2010})}\BibitemShut
  {NoStop}%
\bibitem [{\citenamefont {Singh}\ \emph {et~al.}(2012)\citenamefont {Singh},
  \citenamefont {Manni}, \citenamefont {Reuther}, \citenamefont {Berlijn},
  \citenamefont {Thomale}, \citenamefont {Ku}, \citenamefont {Trebst},\ and\
  \citenamefont {Gegenwart}}]{Singh12}%
  \BibitemOpen
  \bibfield  {author} {\bibinfo {author} {\bibfnamefont {Y.}~\bibnamefont
  {Singh}}, \bibinfo {author} {\bibfnamefont {S.}~\bibnamefont {Manni}},
  \bibinfo {author} {\bibfnamefont {J.}~\bibnamefont {Reuther}}, \bibinfo
  {author} {\bibfnamefont {T.}~\bibnamefont {Berlijn}}, \bibinfo {author}
  {\bibfnamefont {R.}~\bibnamefont {Thomale}}, \bibinfo {author} {\bibfnamefont
  {W.}~\bibnamefont {Ku}}, \bibinfo {author} {\bibfnamefont {S.}~\bibnamefont
  {Trebst}}, \ and\ \bibinfo {author} {\bibfnamefont {P.}~\bibnamefont
  {Gegenwart}},\ }\href {\doibase 10.1103/PhysRevLett.108.127203} {\bibfield
  {journal} {\bibinfo  {journal} {Phys. Rev. Lett.}\ }\textbf {\bibinfo
  {volume} {108}},\ \bibinfo {pages} {127203} (\bibinfo {year}
  {2012})}\BibitemShut {NoStop}%
\bibitem [{\citenamefont {Biffin}\ \emph {et~al.}(2014)\citenamefont {Biffin},
  \citenamefont {Johnson}, \citenamefont {Choi}, \citenamefont {Freund},
  \citenamefont {Manni}, \citenamefont {Bombardi}, \citenamefont {Manuel},
  \citenamefont {Gegenwart},\ and\ \citenamefont {Coldea}}]{Biffin14}%
  \BibitemOpen
  \bibfield  {author} {\bibinfo {author} {\bibfnamefont {A.}~\bibnamefont
  {Biffin}}, \bibinfo {author} {\bibfnamefont {R.~D.}\ \bibnamefont {Johnson}},
  \bibinfo {author} {\bibfnamefont {S.}~\bibnamefont {Choi}}, \bibinfo {author}
  {\bibfnamefont {F.}~\bibnamefont {Freund}}, \bibinfo {author} {\bibfnamefont
  {S.}~\bibnamefont {Manni}}, \bibinfo {author} {\bibfnamefont
  {A.}~\bibnamefont {Bombardi}}, \bibinfo {author} {\bibfnamefont
  {P.}~\bibnamefont {Manuel}}, \bibinfo {author} {\bibfnamefont
  {P.}~\bibnamefont {Gegenwart}}, \ and\ \bibinfo {author} {\bibfnamefont
  {R.}~\bibnamefont {Coldea}},\ }\href {\doibase 10.1103/PhysRevB.90.205116}
  {\bibfield  {journal} {\bibinfo  {journal} {Phys. Rev. B}\ }\textbf {\bibinfo
  {volume} {90}},\ \bibinfo {pages} {205116} (\bibinfo {year}
  {2014})}\BibitemShut {NoStop}%
\bibitem [{\citenamefont {Takayama}\ \emph {et~al.}(2015)\citenamefont
  {Takayama}, \citenamefont {Kato}, \citenamefont {Dinnebier}, \citenamefont
  {Nuss}, \citenamefont {Kono}, \citenamefont {Veiga}, \citenamefont {Fabbris},
  \citenamefont {Haskel},\ and\ \citenamefont {Takagi}}]{Takayama15}%
  \BibitemOpen
  \bibfield  {author} {\bibinfo {author} {\bibfnamefont {T.}~\bibnamefont
  {Takayama}}, \bibinfo {author} {\bibfnamefont {A.}~\bibnamefont {Kato}},
  \bibinfo {author} {\bibfnamefont {R.}~\bibnamefont {Dinnebier}}, \bibinfo
  {author} {\bibfnamefont {J.}~\bibnamefont {Nuss}}, \bibinfo {author}
  {\bibfnamefont {H.}~\bibnamefont {Kono}}, \bibinfo {author} {\bibfnamefont
  {L.~S.~I.}\ \bibnamefont {Veiga}}, \bibinfo {author} {\bibfnamefont
  {G.}~\bibnamefont {Fabbris}}, \bibinfo {author} {\bibfnamefont
  {D.}~\bibnamefont {Haskel}}, \ and\ \bibinfo {author} {\bibfnamefont
  {H.}~\bibnamefont {Takagi}},\ }\href {\doibase
  10.1103/PhysRevLett.114.077202} {\bibfield  {journal} {\bibinfo  {journal}
  {Phys. Rev. Lett.}\ }\textbf {\bibinfo {volume} {114}},\ \bibinfo {pages}
  {077202} (\bibinfo {year} {2015})}\BibitemShut {NoStop}%
\bibitem [{\citenamefont {Modic}\ \emph {et~al.}(2014)\citenamefont {Modic},
  \citenamefont {Smidt}, \citenamefont {Kimchi}, \citenamefont {Breznay},
  \citenamefont {Biffin}, \citenamefont {Choi}, \citenamefont {Johnson},
  \citenamefont {Coldea}, \citenamefont {Watkins-Curry}, \citenamefont
  {McCandless}, \citenamefont {Chan}, \citenamefont {Gandara}, \citenamefont
  {Islam}, \citenamefont {Vishwanath}, \citenamefont {Shekhter}, \citenamefont
  {McDonald},\ and\ \citenamefont {Analytis}}]{Modic14}%
  \BibitemOpen
  \bibfield  {author} {\bibinfo {author} {\bibfnamefont {K.~A.}\ \bibnamefont
  {Modic}}, \bibinfo {author} {\bibfnamefont {T.~E.}\ \bibnamefont {Smidt}},
  \bibinfo {author} {\bibfnamefont {I.}~\bibnamefont {Kimchi}}, \bibinfo
  {author} {\bibfnamefont {N.~P.}\ \bibnamefont {Breznay}}, \bibinfo {author}
  {\bibfnamefont {A.}~\bibnamefont {Biffin}}, \bibinfo {author} {\bibfnamefont
  {S.}~\bibnamefont {Choi}}, \bibinfo {author} {\bibfnamefont {R.~D.}\
  \bibnamefont {Johnson}}, \bibinfo {author} {\bibfnamefont {R.}~\bibnamefont
  {Coldea}}, \bibinfo {author} {\bibfnamefont {P.}~\bibnamefont
  {Watkins-Curry}}, \bibinfo {author} {\bibfnamefont {G.~T.}\ \bibnamefont
  {McCandless}}, \bibinfo {author} {\bibfnamefont {J.~Y.}\ \bibnamefont
  {Chan}}, \bibinfo {author} {\bibfnamefont {F.}~\bibnamefont {Gandara}},
  \bibinfo {author} {\bibfnamefont {Z.}~\bibnamefont {Islam}}, \bibinfo
  {author} {\bibfnamefont {A.}~\bibnamefont {Vishwanath}}, \bibinfo {author}
  {\bibfnamefont {A.}~\bibnamefont {Shekhter}}, \bibinfo {author}
  {\bibfnamefont {R.~D.}\ \bibnamefont {McDonald}}, \ and\ \bibinfo {author}
  {\bibfnamefont {J.~G.}\ \bibnamefont {Analytis}},\ }\href {\doibase
  10.1038/ncomms5203} {\bibfield  {journal} {\bibinfo  {journal} {Nat.
  Commun.}\ }\textbf {\bibinfo {volume} {5}},\ \bibinfo {pages} {4203}
  (\bibinfo {year} {2014})}\BibitemShut {NoStop}%
\bibitem [{\citenamefont {Okamoto}\ \emph {et~al.}(2007)\citenamefont
  {Okamoto}, \citenamefont {Nohara}, \citenamefont {Aruga-Katori},\ and\
  \citenamefont {Takagi}}]{Okamoto07}%
  \BibitemOpen
  \bibfield  {author} {\bibinfo {author} {\bibfnamefont {Y.}~\bibnamefont
  {Okamoto}}, \bibinfo {author} {\bibfnamefont {M.}~\bibnamefont {Nohara}},
  \bibinfo {author} {\bibfnamefont {H.}~\bibnamefont {Aruga-Katori}}, \ and\
  \bibinfo {author} {\bibfnamefont {H.}~\bibnamefont {Takagi}},\ }\href
  {\doibase 10.1103/PhysRevLett.99.137207} {\bibfield  {journal} {\bibinfo
  {journal} {Phys. Rev. Lett.}\ }\textbf {\bibinfo {volume} {99}},\ \bibinfo
  {pages} {137207} (\bibinfo {year} {2007})}\BibitemShut {NoStop}%
\bibitem [{\citenamefont {Jackeli}\ and\ \citenamefont
  {Khaliullin}(2009)}]{Jackeli09}%
  \BibitemOpen
  \bibfield  {author} {\bibinfo {author} {\bibfnamefont {G.}~\bibnamefont
  {Jackeli}}\ and\ \bibinfo {author} {\bibfnamefont {G.}~\bibnamefont
  {Khaliullin}},\ }\href@noop {} {\bibfield  {journal} {\bibinfo  {journal}
  {Phys. Rev. Lett.}\ }\textbf {\bibinfo {volume} {102}},\ \bibinfo {pages}
  {017205} (\bibinfo {year} {2009})}\BibitemShut {NoStop}%
\bibitem [{\citenamefont {Chaloupka}\ \emph {et~al.}(2010)\citenamefont
  {Chaloupka}, \citenamefont {Jackeli},\ and\ \citenamefont
  {Khaliullin}}]{Chaloupka10}%
  \BibitemOpen
  \bibfield  {author} {\bibinfo {author} {\bibfnamefont {J.}~\bibnamefont
  {Chaloupka}}, \bibinfo {author} {\bibfnamefont {G.}~\bibnamefont {Jackeli}},
  \ and\ \bibinfo {author} {\bibfnamefont {G.}~\bibnamefont {Khaliullin}},\
  }\href {\doibase 10.1103/PhysRevLett.105.027204} {\bibfield  {journal}
  {\bibinfo  {journal} {Phys. Rev. Lett.}\ }\textbf {\bibinfo {volume} {105}},\
  \bibinfo {pages} {027204} (\bibinfo {year} {2010})}\BibitemShut {NoStop}%
\bibitem [{\citenamefont {Chaloupka}\ \emph {et~al.}(2013)\citenamefont
  {Chaloupka}, \citenamefont {Jackeli},\ and\ \citenamefont
  {Khaliullin}}]{Chaloupka13}%
  \BibitemOpen
  \bibfield  {author} {\bibinfo {author} {\bibfnamefont {J.}~\bibnamefont
  {Chaloupka}}, \bibinfo {author} {\bibfnamefont {G.}~\bibnamefont {Jackeli}},
  \ and\ \bibinfo {author} {\bibfnamefont {G.}~\bibnamefont {Khaliullin}},\
  }\href {\doibase 10.1103/PhysRevLett.110.097204} {\bibfield  {journal}
  {\bibinfo  {journal} {Phys. Rev. Lett.}\ }\textbf {\bibinfo {volume} {110}},\
  \bibinfo {pages} {097204} (\bibinfo {year} {2013})}\BibitemShut {NoStop}%
\bibitem [{\citenamefont {Khaliullin}(2013)}]{Khaliullin_2013}%
  \BibitemOpen
  \bibfield  {author} {\bibinfo {author} {\bibfnamefont {G.}~\bibnamefont
  {Khaliullin}},\ }\href {\doibase 10.1103/physrevlett.111.197201} {\bibfield
  {journal} {\bibinfo  {journal} {Phys. Rev. Lett.}\ }\textbf {\bibinfo
  {volume} {111}},\ \bibinfo {pages} {197201} (\bibinfo {year}
  {2013})}\BibitemShut {NoStop}%
\bibitem [{\citenamefont {Cao}\ \emph {et~al.}(2014)\citenamefont {Cao},
  \citenamefont {Qi}, \citenamefont {Li}, \citenamefont {Terzic}, \citenamefont
  {Yuan}, \citenamefont {DeLong}, \citenamefont {Murthy},\ and\ \citenamefont
  {Kaul}}]{Cao_2014}%
  \BibitemOpen
  \bibfield  {author} {\bibinfo {author} {\bibfnamefont {G.}~\bibnamefont
  {Cao}}, \bibinfo {author} {\bibfnamefont {T.~F.}\ \bibnamefont {Qi}},
  \bibinfo {author} {\bibfnamefont {L.}~\bibnamefont {Li}}, \bibinfo {author}
  {\bibfnamefont {J.}~\bibnamefont {Terzic}}, \bibinfo {author} {\bibfnamefont
  {S.~J.}\ \bibnamefont {Yuan}}, \bibinfo {author} {\bibfnamefont {L.~E.}\
  \bibnamefont {DeLong}}, \bibinfo {author} {\bibfnamefont {G.}~\bibnamefont
  {Murthy}}, \ and\ \bibinfo {author} {\bibfnamefont {R.~K.}\ \bibnamefont
  {Kaul}},\ }\href {\doibase 10.1103/physrevlett.112.056402} {\bibfield
  {journal} {\bibinfo  {journal} {Phys. Rev. Lett.}\ }\textbf {\bibinfo
  {volume} {112}},\ \bibinfo {pages} {056402} (\bibinfo {year}
  {2014})}\BibitemShut {NoStop}%
\bibitem [{\citenamefont {Terzic}\ \emph {et~al.}(2017)\citenamefont {Terzic},
  \citenamefont {Zheng}, \citenamefont {Ye}, \citenamefont {Zhao},
  \citenamefont {Schlottmann}, \citenamefont {De~Long}, \citenamefont {Yuan},\
  and\ \citenamefont {Cao}}]{Terzic17}%
  \BibitemOpen
  \bibfield  {author} {\bibinfo {author} {\bibfnamefont {J.}~\bibnamefont
  {Terzic}}, \bibinfo {author} {\bibfnamefont {H.}~\bibnamefont {Zheng}},
  \bibinfo {author} {\bibfnamefont {F.}~\bibnamefont {Ye}}, \bibinfo {author}
  {\bibfnamefont {H.~D.}\ \bibnamefont {Zhao}}, \bibinfo {author}
  {\bibfnamefont {P.}~\bibnamefont {Schlottmann}}, \bibinfo {author}
  {\bibfnamefont {L.~E.}\ \bibnamefont {De~Long}}, \bibinfo {author}
  {\bibfnamefont {S.~J.}\ \bibnamefont {Yuan}}, \ and\ \bibinfo {author}
  {\bibfnamefont {G.}~\bibnamefont {Cao}},\ }\href {\doibase
  10.1103/PhysRevB.96.064436} {\bibfield  {journal} {\bibinfo  {journal} {Phys.
  Rev. B}\ }\textbf {\bibinfo {volume} {96}},\ \bibinfo {pages} {064436}
  (\bibinfo {year} {2017})}\BibitemShut {NoStop}%
\bibitem [{\citenamefont {Ranjbar}\ \emph {et~al.}(2015)\citenamefont
  {Ranjbar}, \citenamefont {Reynolds}, \citenamefont {Kayser}, \citenamefont
  {Kennedy}, \citenamefont {Hester},\ and\ \citenamefont
  {Kimpton}}]{Ranjbar15}%
  \BibitemOpen
  \bibfield  {author} {\bibinfo {author} {\bibfnamefont {B.}~\bibnamefont
  {Ranjbar}}, \bibinfo {author} {\bibfnamefont {E.}~\bibnamefont {Reynolds}},
  \bibinfo {author} {\bibfnamefont {P.}~\bibnamefont {Kayser}}, \bibinfo
  {author} {\bibfnamefont {B.~J.}\ \bibnamefont {Kennedy}}, \bibinfo {author}
  {\bibfnamefont {J.~R.}\ \bibnamefont {Hester}}, \ and\ \bibinfo {author}
  {\bibfnamefont {J.~A.}\ \bibnamefont {Kimpton}},\ }\href {\doibase
  10.1021/acs.inorgchem.5b01905} {\bibfield  {journal} {\bibinfo  {journal}
  {Inorg. Chem.}\ }\textbf {\bibinfo {volume} {54}},\ \bibinfo {pages} {10468}
  (\bibinfo {year} {2015})}\BibitemShut {NoStop}%
\bibitem [{\citenamefont {Dey}\ \emph {et~al.}(2016)\citenamefont {Dey},
  \citenamefont {Maljuk}, \citenamefont {Efremov}, \citenamefont {Kataeva},
  \citenamefont {Gass}, \citenamefont {Blum}, \citenamefont {Steckel},
  \citenamefont {Gruner}, \citenamefont {Ritschel}, \citenamefont {Wolter},\
  and\ \citenamefont {et~al.}}]{Dey_2016}%
  \BibitemOpen
  \bibfield  {author} {\bibinfo {author} {\bibfnamefont {T.}~\bibnamefont
  {Dey}}, \bibinfo {author} {\bibfnamefont {A.}~\bibnamefont {Maljuk}},
  \bibinfo {author} {\bibfnamefont {D.~V.}\ \bibnamefont {Efremov}}, \bibinfo
  {author} {\bibfnamefont {O.}~\bibnamefont {Kataeva}}, \bibinfo {author}
  {\bibfnamefont {S.}~\bibnamefont {Gass}}, \bibinfo {author} {\bibfnamefont
  {C.~G.~F.}\ \bibnamefont {Blum}}, \bibinfo {author} {\bibfnamefont
  {F.}~\bibnamefont {Steckel}}, \bibinfo {author} {\bibfnamefont
  {D.}~\bibnamefont {Gruner}}, \bibinfo {author} {\bibfnamefont
  {T.}~\bibnamefont {Ritschel}}, \bibinfo {author} {\bibfnamefont {A.~U.~B.}\
  \bibnamefont {Wolter}}, \ and\ \bibinfo {author} {\bibnamefont {et~al.}},\
  }\href {\doibase 10.1103/physrevb.93.014434} {\bibfield  {journal} {\bibinfo
  {journal} {Phys. Rev. B}\ }\textbf {\bibinfo {volume} {93}},\ \bibinfo
  {pages} {014434} (\bibinfo {year} {2016})}\BibitemShut {NoStop}%
\bibitem [{\citenamefont {Phelan}\ \emph {et~al.}(2016)\citenamefont {Phelan},
  \citenamefont {Seibel}, \citenamefont {Badoe}, \citenamefont {Xie},\ and\
  \citenamefont {Cava}}]{Phelan_2016}%
  \BibitemOpen
  \bibfield  {author} {\bibinfo {author} {\bibfnamefont {B.~F.}\ \bibnamefont
  {Phelan}}, \bibinfo {author} {\bibfnamefont {E.~M.}\ \bibnamefont {Seibel}},
  \bibinfo {author} {\bibfnamefont {D.}~\bibnamefont {Badoe}}, \bibinfo
  {author} {\bibfnamefont {W.}~\bibnamefont {Xie}}, \ and\ \bibinfo {author}
  {\bibfnamefont {R.}~\bibnamefont {Cava}},\ }\href {\doibase
  10.1016/j.ssc.2016.03.017} {\bibfield  {journal} {\bibinfo  {journal} {Solid
  State Commun.}\ }\textbf {\bibinfo {volume} {236}},\ \bibinfo {pages} {37}
  (\bibinfo {year} {2016})}\BibitemShut {NoStop}%
\bibitem [{\citenamefont {Chen}\ \emph {et~al.}(2017)\citenamefont {Chen},
  \citenamefont {Svoboda}, \citenamefont {Zheng}, \citenamefont {Sales},
  \citenamefont {Mandrus}, \citenamefont {Zhou}, \citenamefont {Zhou},
  \citenamefont {McComb}, \citenamefont {Randeria}, \citenamefont {Trivedi},\
  and\ \citenamefont {Yan}}]{Chen17}%
  \BibitemOpen
  \bibfield  {author} {\bibinfo {author} {\bibfnamefont {Q.}~\bibnamefont
  {Chen}}, \bibinfo {author} {\bibfnamefont {C.}~\bibnamefont {Svoboda}},
  \bibinfo {author} {\bibfnamefont {Q.}~\bibnamefont {Zheng}}, \bibinfo
  {author} {\bibfnamefont {B.~C.}\ \bibnamefont {Sales}}, \bibinfo {author}
  {\bibfnamefont {D.~G.}\ \bibnamefont {Mandrus}}, \bibinfo {author}
  {\bibfnamefont {H.~D.}\ \bibnamefont {Zhou}}, \bibinfo {author}
  {\bibfnamefont {J.-S.}\ \bibnamefont {Zhou}}, \bibinfo {author}
  {\bibfnamefont {D.}~\bibnamefont {McComb}}, \bibinfo {author} {\bibfnamefont
  {M.}~\bibnamefont {Randeria}}, \bibinfo {author} {\bibfnamefont
  {N.}~\bibnamefont {Trivedi}}, \ and\ \bibinfo {author} {\bibfnamefont
  {J.-Q.}\ \bibnamefont {Yan}},\ }\href {\doibase 10.1103/PhysRevB.96.144423}
  {\bibfield  {journal} {\bibinfo  {journal} {Phys. Rev. B}\ }\textbf {\bibinfo
  {volume} {96}},\ \bibinfo {pages} {144423} (\bibinfo {year}
  {2017})}\BibitemShut {NoStop}%
\bibitem [{\citenamefont {Corredor}\ \emph {et~al.}(2017)\citenamefont
  {Corredor}, \citenamefont {Aslan-Cansever}, \citenamefont {Sturza},
  \citenamefont {Manna}, \citenamefont {Maljuk}, \citenamefont {Gass},
  \citenamefont {Dey}, \citenamefont {Wolter}, \citenamefont {Kataeva},
  \citenamefont {Zimmermann}, \citenamefont {Geyer}, \citenamefont {Blum},
  \citenamefont {Wurmehl},\ and\ \citenamefont {B\"uchner}}]{Corredor17}%
  \BibitemOpen
  \bibfield  {author} {\bibinfo {author} {\bibfnamefont {L.~T.}\ \bibnamefont
  {Corredor}}, \bibinfo {author} {\bibfnamefont {G.}~\bibnamefont
  {Aslan-Cansever}}, \bibinfo {author} {\bibfnamefont {M.}~\bibnamefont
  {Sturza}}, \bibinfo {author} {\bibfnamefont {K.}~\bibnamefont {Manna}},
  \bibinfo {author} {\bibfnamefont {A.}~\bibnamefont {Maljuk}}, \bibinfo
  {author} {\bibfnamefont {S.}~\bibnamefont {Gass}}, \bibinfo {author}
  {\bibfnamefont {T.}~\bibnamefont {Dey}}, \bibinfo {author} {\bibfnamefont
  {A.~U.~B.}\ \bibnamefont {Wolter}}, \bibinfo {author} {\bibfnamefont
  {O.}~\bibnamefont {Kataeva}}, \bibinfo {author} {\bibfnamefont
  {A.}~\bibnamefont {Zimmermann}}, \bibinfo {author} {\bibfnamefont
  {M.}~\bibnamefont {Geyer}}, \bibinfo {author} {\bibfnamefont {C.~G.~F.}\
  \bibnamefont {Blum}}, \bibinfo {author} {\bibfnamefont {S.}~\bibnamefont
  {Wurmehl}}, \ and\ \bibinfo {author} {\bibfnamefont {B.}~\bibnamefont
  {B\"uchner}},\ }\href {\doibase 10.1103/PhysRevB.95.064418} {\bibfield
  {journal} {\bibinfo  {journal} {Phys. Rev. B}\ }\textbf {\bibinfo {volume}
  {95}},\ \bibinfo {pages} {064418} (\bibinfo {year} {2017})}\BibitemShut
  {NoStop}%
\bibitem [{\citenamefont {Hammerath}\ \emph {et~al.}(2017)\citenamefont
  {Hammerath}, \citenamefont {Sarkar}, \citenamefont {Kamusella}, \citenamefont
  {Baines}, \citenamefont {Klauss}, \citenamefont {Dey}, \citenamefont
  {Maljuk}, \citenamefont {Ga\ss{}}, \citenamefont {Wolter}, \citenamefont
  {Grafe}, \citenamefont {Wurmehl},\ and\ \citenamefont
  {B\"uchner}}]{Hammerath17}%
  \BibitemOpen
  \bibfield  {author} {\bibinfo {author} {\bibfnamefont {F.}~\bibnamefont
  {Hammerath}}, \bibinfo {author} {\bibfnamefont {R.}~\bibnamefont {Sarkar}},
  \bibinfo {author} {\bibfnamefont {S.}~\bibnamefont {Kamusella}}, \bibinfo
  {author} {\bibfnamefont {C.}~\bibnamefont {Baines}}, \bibinfo {author}
  {\bibfnamefont {H.-H.}\ \bibnamefont {Klauss}}, \bibinfo {author}
  {\bibfnamefont {T.}~\bibnamefont {Dey}}, \bibinfo {author} {\bibfnamefont
  {A.}~\bibnamefont {Maljuk}}, \bibinfo {author} {\bibfnamefont
  {S.}~\bibnamefont {Ga\ss{}}}, \bibinfo {author} {\bibfnamefont {A.~U.~B.}\
  \bibnamefont {Wolter}}, \bibinfo {author} {\bibfnamefont {H.-J.}\
  \bibnamefont {Grafe}}, \bibinfo {author} {\bibfnamefont {S.}~\bibnamefont
  {Wurmehl}}, \ and\ \bibinfo {author} {\bibfnamefont {B.}~\bibnamefont
  {B\"uchner}},\ }\href {\doibase 10.1103/PhysRevB.96.165108} {\bibfield
  {journal} {\bibinfo  {journal} {Phys. Rev. B}\ }\textbf {\bibinfo {volume}
  {96}},\ \bibinfo {pages} {165108} (\bibinfo {year} {2017})}\BibitemShut
  {NoStop}%
\bibitem [{\citenamefont {Bhowal}\ \emph {et~al.}(2015)\citenamefont {Bhowal},
  \citenamefont {Baidya}, \citenamefont {Dasgupta},\ and\ \citenamefont
  {Saha-Dasgupta}}]{Bhowal15}%
  \BibitemOpen
  \bibfield  {author} {\bibinfo {author} {\bibfnamefont {S.}~\bibnamefont
  {Bhowal}}, \bibinfo {author} {\bibfnamefont {S.}~\bibnamefont {Baidya}},
  \bibinfo {author} {\bibfnamefont {I.}~\bibnamefont {Dasgupta}}, \ and\
  \bibinfo {author} {\bibfnamefont {T.}~\bibnamefont {Saha-Dasgupta}},\ }\href
  {\doibase 10.1103/PhysRevB.92.121113} {\bibfield  {journal} {\bibinfo
  {journal} {Phys. Rev. B}\ }\textbf {\bibinfo {volume} {92}},\ \bibinfo
  {pages} {121113} (\bibinfo {year} {2015})}\BibitemShut {NoStop}%
\bibitem [{\citenamefont {Meetei}\ \emph {et~al.}(2015)\citenamefont {Meetei},
  \citenamefont {Cole}, \citenamefont {Randeria},\ and\ \citenamefont
  {Trivedi}}]{Meetei15}%
  \BibitemOpen
  \bibfield  {author} {\bibinfo {author} {\bibfnamefont {O.~N.}\ \bibnamefont
  {Meetei}}, \bibinfo {author} {\bibfnamefont {W.~S.}\ \bibnamefont {Cole}},
  \bibinfo {author} {\bibfnamefont {M.}~\bibnamefont {Randeria}}, \ and\
  \bibinfo {author} {\bibfnamefont {N.}~\bibnamefont {Trivedi}},\ }\href
  {\doibase 10.1103/PhysRevB.91.054412} {\bibfield  {journal} {\bibinfo
  {journal} {Phys. Rev. B}\ }\textbf {\bibinfo {volume} {91}},\ \bibinfo
  {pages} {054412} (\bibinfo {year} {2015})}\BibitemShut {NoStop}%
\bibitem [{\citenamefont {Pajskr}\ \emph {et~al.}(2016)\citenamefont {Pajskr},
  \citenamefont {Nov\'ak}, \citenamefont {Pokorn\'y}, \citenamefont
  {Koloren\ifmmode~\check{c}\else \v{c}\fi{}}, \citenamefont {Arita},\ and\
  \citenamefont {Kune\ifmmode~\check{s}\else \v{s}\fi{}}}]{Pajskr16}%
  \BibitemOpen
  \bibfield  {author} {\bibinfo {author} {\bibfnamefont {K.}~\bibnamefont
  {Pajskr}}, \bibinfo {author} {\bibfnamefont {P.}~\bibnamefont {Nov\'ak}},
  \bibinfo {author} {\bibfnamefont {V.}~\bibnamefont {Pokorn\'y}}, \bibinfo
  {author} {\bibfnamefont {J.}~\bibnamefont {Koloren\ifmmode~\check{c}\else
  \v{c}\fi{}}}, \bibinfo {author} {\bibfnamefont {R.}~\bibnamefont {Arita}}, \
  and\ \bibinfo {author} {\bibfnamefont {J.}~\bibnamefont
  {Kune\ifmmode~\check{s}\else \v{s}\fi{}}},\ }\href {\doibase
  10.1103/PhysRevB.93.035129} {\bibfield  {journal} {\bibinfo  {journal} {Phys.
  Rev. B}\ }\textbf {\bibinfo {volume} {93}},\ \bibinfo {pages} {035129}
  (\bibinfo {year} {2016})}\BibitemShut {NoStop}%
\bibitem [{\citenamefont {Nag}\ \emph {et~al.}(2016)\citenamefont {Nag},
  \citenamefont {Middey}, \citenamefont {Bhowal}, \citenamefont {Panda},
  \citenamefont {Mathieu}, \citenamefont {Orain}, \citenamefont {Bert},
  \citenamefont {Mendels}, \citenamefont {Freeman}, \citenamefont {Mansson},
  \citenamefont {Ronnow}, \citenamefont {Telling}, \citenamefont {Biswas},
  \citenamefont {Sheptyakov}, \citenamefont {Kaushik}, \citenamefont
  {Siruguri}, \citenamefont {Meneghini}, \citenamefont {Sarma}, \citenamefont
  {Dasgupta},\ and\ \citenamefont {Ray}}]{Nag16}%
  \BibitemOpen
  \bibfield  {author} {\bibinfo {author} {\bibfnamefont {A.}~\bibnamefont
  {Nag}}, \bibinfo {author} {\bibfnamefont {S.}~\bibnamefont {Middey}},
  \bibinfo {author} {\bibfnamefont {S.}~\bibnamefont {Bhowal}}, \bibinfo
  {author} {\bibfnamefont {S.~K.}\ \bibnamefont {Panda}}, \bibinfo {author}
  {\bibfnamefont {R.}~\bibnamefont {Mathieu}}, \bibinfo {author} {\bibfnamefont
  {J.~C.}\ \bibnamefont {Orain}}, \bibinfo {author} {\bibfnamefont
  {F.}~\bibnamefont {Bert}}, \bibinfo {author} {\bibfnamefont {P.}~\bibnamefont
  {Mendels}}, \bibinfo {author} {\bibfnamefont {P.~G.}\ \bibnamefont
  {Freeman}}, \bibinfo {author} {\bibfnamefont {M.}~\bibnamefont {Mansson}},
  \bibinfo {author} {\bibfnamefont {H.~M.}\ \bibnamefont {Ronnow}}, \bibinfo
  {author} {\bibfnamefont {M.}~\bibnamefont {Telling}}, \bibinfo {author}
  {\bibfnamefont {P.~K.}\ \bibnamefont {Biswas}}, \bibinfo {author}
  {\bibfnamefont {D.}~\bibnamefont {Sheptyakov}}, \bibinfo {author}
  {\bibfnamefont {S.~D.}\ \bibnamefont {Kaushik}}, \bibinfo {author}
  {\bibfnamefont {V.}~\bibnamefont {Siruguri}}, \bibinfo {author}
  {\bibfnamefont {C.}~\bibnamefont {Meneghini}}, \bibinfo {author}
  {\bibfnamefont {D.~D.}\ \bibnamefont {Sarma}}, \bibinfo {author}
  {\bibfnamefont {I.}~\bibnamefont {Dasgupta}}, \ and\ \bibinfo {author}
  {\bibfnamefont {S.}~\bibnamefont {Ray}},\ }\href {\doibase
  10.1103/PhysRevLett.116.097205} {\bibfield  {journal} {\bibinfo  {journal}
  {Phys. Rev. Lett.}\ }\textbf {\bibinfo {volume} {116}},\ \bibinfo {pages}
  {097205} (\bibinfo {year} {2016})}\BibitemShut {NoStop}%
\bibitem [{\citenamefont {Kim}\ \emph {et~al.}(2017)\citenamefont {Kim},
  \citenamefont {Jeschke}, \citenamefont {Werner},\ and\ \citenamefont
  {Valent\'{\i}}}]{Kim17}%
  \BibitemOpen
  \bibfield  {author} {\bibinfo {author} {\bibfnamefont {A.~J.}\ \bibnamefont
  {Kim}}, \bibinfo {author} {\bibfnamefont {H.~O.}\ \bibnamefont {Jeschke}},
  \bibinfo {author} {\bibfnamefont {P.}~\bibnamefont {Werner}}, \ and\ \bibinfo
  {author} {\bibfnamefont {R.}~\bibnamefont {Valent\'{\i}}},\ }\href {\doibase
  10.1103/PhysRevLett.118.086401} {\bibfield  {journal} {\bibinfo  {journal}
  {Phys. Rev. Lett.}\ }\textbf {\bibinfo {volume} {118}},\ \bibinfo {pages}
  {086401} (\bibinfo {year} {2017})}\BibitemShut {NoStop}%
\bibitem [{\citenamefont {Svoboda}\ \emph {et~al.}(2017)\citenamefont
  {Svoboda}, \citenamefont {Randeria},\ and\ \citenamefont
  {Trivedi}}]{Svoboda17}%
  \BibitemOpen
  \bibfield  {author} {\bibinfo {author} {\bibfnamefont {C.}~\bibnamefont
  {Svoboda}}, \bibinfo {author} {\bibfnamefont {M.}~\bibnamefont {Randeria}}, \
  and\ \bibinfo {author} {\bibfnamefont {N.}~\bibnamefont {Trivedi}},\ }\href
  {\doibase 10.1103/PhysRevB.95.014409} {\bibfield  {journal} {\bibinfo
  {journal} {Phys. Rev. B}\ }\textbf {\bibinfo {volume} {95}},\ \bibinfo
  {pages} {014409} (\bibinfo {year} {2017})}\BibitemShut {NoStop}%
\bibitem [{\citenamefont {Golze}\ \emph {et~al.}(2006)\citenamefont {Golze},
  \citenamefont {Alfonsov}, \citenamefont {Klingeler}, \citenamefont
  {B{\"u}chner}, \citenamefont {Kataev}, \citenamefont {Mennerich},
  \citenamefont {Klauss}, \citenamefont {Goiran}, \citenamefont {Broto},
  \citenamefont {Rakoto},\ and\ \citenamefont {et~al.}}]{Golze_2006}%
  \BibitemOpen
  \bibfield  {author} {\bibinfo {author} {\bibfnamefont {C.}~\bibnamefont
  {Golze}}, \bibinfo {author} {\bibfnamefont {A.}~\bibnamefont {Alfonsov}},
  \bibinfo {author} {\bibfnamefont {R.}~\bibnamefont {Klingeler}}, \bibinfo
  {author} {\bibfnamefont {B.}~\bibnamefont {B{\"u}chner}}, \bibinfo {author}
  {\bibfnamefont {V.}~\bibnamefont {Kataev}}, \bibinfo {author} {\bibfnamefont
  {C.}~\bibnamefont {Mennerich}}, \bibinfo {author} {\bibfnamefont {H.-H.}\
  \bibnamefont {Klauss}}, \bibinfo {author} {\bibfnamefont {M.}~\bibnamefont
  {Goiran}}, \bibinfo {author} {\bibfnamefont {J.-M.}\ \bibnamefont {Broto}},
  \bibinfo {author} {\bibfnamefont {H.}~\bibnamefont {Rakoto}}, \ and\ \bibinfo
  {author} {\bibnamefont {et~al.}},\ }\href {\doibase
  10.1103/physrevb.73.224403} {\bibfield  {journal} {\bibinfo  {journal} {Phys.
  Rev. B}\ }\textbf {\bibinfo {volume} {73}},\ \bibinfo {pages} {224403}
  (\bibinfo {year} {2006})}\BibitemShut {NoStop}%
\bibitem [{\citenamefont {Abragam}\ and\ \citenamefont
  {Bleaney}(2012)}]{abragam}%
  \BibitemOpen
  \bibfield  {author} {\bibinfo {author} {\bibfnamefont {A.}~\bibnamefont
  {Abragam}}\ and\ \bibinfo {author} {\bibfnamefont {B.}~\bibnamefont
  {Bleaney}},\ }\href {https://books.google.se/books?id=SSD7AAAAQBAJ} {\emph
  {\bibinfo {title} {Electron Paramagnetic Resonance of Transition Ions}}},\
  International Series of Monographs on Physics\ (\bibinfo  {publisher} {Oxford
  university, New York},\ \bibinfo {year} {2012})\BibitemShut {NoStop}%
\bibitem [{\citenamefont {Chang}\ \emph {et~al.}(1978)\citenamefont {Chang},
  \citenamefont {Foster},\ and\ \citenamefont {Kahn}}]{Rubin1}%
  \BibitemOpen
  \bibfield  {author} {\bibinfo {author} {\bibfnamefont {T.}~\bibnamefont
  {Chang}}, \bibinfo {author} {\bibfnamefont {D.}~\bibnamefont {Foster}}, \
  and\ \bibinfo {author} {\bibfnamefont {A.}~\bibnamefont {Kahn}},\ }\href
  {\doibase 10.6028/jres.083.008} {\bibfield  {journal} {\bibinfo  {journal}
  {J. Res. Bur. Stand.}\ }\textbf {\bibinfo {volume} {83}},\ \bibinfo {pages}
  {133} (\bibinfo {year} {1978})}\BibitemShut {NoStop}%
\bibitem [{\citenamefont {Chang}\ and\ \citenamefont {Kahn}(1978)}]{Rubin2}%
  \BibitemOpen
  \bibfield  {author} {\bibinfo {author} {\bibfnamefont {T.}~\bibnamefont
  {Chang}}\ and\ \bibinfo {author} {\bibfnamefont {A.~H.}\ \bibnamefont
  {Kahn}},\ }\href@noop {} {\emph {\bibinfo {title} {Standard Reference
  Materials : Electron Paramagnetic Resonance Intensity Standard : SRM 2601;
  description and use}}},\ \bibinfo {series} {NBS/NIST Special Publications},
  Vol.\ \bibinfo {volume} {26-59}\ (\bibinfo  {publisher} {National Bureau of
  Standards, washington, DC},\ \bibinfo {year} {1978})\BibitemShut {NoStop}%
\bibitem [{\citenamefont {Huber}(1972)}]{Huber72}%
  \BibitemOpen
  \bibfield  {author} {\bibinfo {author} {\bibfnamefont {D.~L.}\ \bibnamefont
  {Huber}},\ }\href {\doibase 10.1103/PhysRevB.6.3180} {\bibfield  {journal}
  {\bibinfo  {journal} {Phys. Rev. B}\ }\textbf {\bibinfo {volume} {6}},\
  \bibinfo {pages} {3180} (\bibinfo {year} {1972})}\BibitemShut {NoStop}%
\bibitem [{\citenamefont {Huber}\ and\ \citenamefont {Seehra}(1975)}]{Huber75}%
  \BibitemOpen
  \bibfield  {author} {\bibinfo {author} {\bibfnamefont {D.~L.}\ \bibnamefont
  {Huber}}\ and\ \bibinfo {author} {\bibfnamefont {M.~S.}\ \bibnamefont
  {Seehra}},\ }\href {\doibase 10.1016/0022-3697(75)90094-3} {\bibfield
  {journal} {\bibinfo  {journal} {J. Phys. Chem. Solids}\ }\textbf {\bibinfo
  {volume} {36}},\ \bibinfo {pages} {723} (\bibinfo {year} {1975})}\BibitemShut
  {NoStop}%
\bibitem [{\citenamefont {Choy}\ \emph {et~al.}(1995)\citenamefont {Choy},
  \citenamefont {Kim}, \citenamefont {Hwang}, \citenamefont {Demazeau},\ and\
  \citenamefont {Jung}}]{Choy_1995}%
  \BibitemOpen
  \bibfield  {author} {\bibinfo {author} {\bibfnamefont {J.-H.}\ \bibnamefont
  {Choy}}, \bibinfo {author} {\bibfnamefont {D.-K.}\ \bibnamefont {Kim}},
  \bibinfo {author} {\bibfnamefont {S.-H.}\ \bibnamefont {Hwang}}, \bibinfo
  {author} {\bibfnamefont {G.}~\bibnamefont {Demazeau}}, \ and\ \bibinfo
  {author} {\bibfnamefont {D.-Y.}\ \bibnamefont {Jung}},\ }\href {\doibase
  10.1021/ja00138a010} {\bibfield  {journal} {\bibinfo  {journal} {J. Am. Chem.
  Soc.}\ }\textbf {\bibinfo {volume} {117}},\ \bibinfo {pages} {8557} (\bibinfo
  {year} {1995})}\BibitemShut {NoStop}%
\bibitem [{\citenamefont {Clancy}\ \emph {et~al.}(2012)\citenamefont {Clancy},
  \citenamefont {Chen}, \citenamefont {Kim}, \citenamefont {Chen},
  \citenamefont {Plumb}, \citenamefont {Jeon}, \citenamefont {Noh},\ and\
  \citenamefont {Kim}}]{Clancy12}%
  \BibitemOpen
  \bibfield  {author} {\bibinfo {author} {\bibfnamefont {J.~P.}\ \bibnamefont
  {Clancy}}, \bibinfo {author} {\bibfnamefont {N.}~\bibnamefont {Chen}},
  \bibinfo {author} {\bibfnamefont {C.~Y.}\ \bibnamefont {Kim}}, \bibinfo
  {author} {\bibfnamefont {W.~F.}\ \bibnamefont {Chen}}, \bibinfo {author}
  {\bibfnamefont {K.~W.}\ \bibnamefont {Plumb}}, \bibinfo {author}
  {\bibfnamefont {B.~C.}\ \bibnamefont {Jeon}}, \bibinfo {author}
  {\bibfnamefont {T.~W.}\ \bibnamefont {Noh}}, \ and\ \bibinfo {author}
  {\bibfnamefont {Y.-J.}\ \bibnamefont {Kim}},\ }\href {\doibase
  10.1103/PhysRevB.86.195131} {\bibfield  {journal} {\bibinfo  {journal} {Phys.
  Rev. B}\ }\textbf {\bibinfo {volume} {86}},\ \bibinfo {pages} {195131}
  (\bibinfo {year} {2012})}\BibitemShut {NoStop}%
\bibitem [{\citenamefont {Allen}\ \emph {et~al.}(1972)\citenamefont {Allen},
  \citenamefont {El-Sharkawy},\ and\ \citenamefont {Warren}}]{Allen72}%
  \BibitemOpen
  \bibfield  {author} {\bibinfo {author} {\bibfnamefont {G.~C.}\ \bibnamefont
  {Allen}}, \bibinfo {author} {\bibfnamefont {G.~A.~M.}\ \bibnamefont
  {El-Sharkawy}}, \ and\ \bibinfo {author} {\bibfnamefont {K.~D.}\ \bibnamefont
  {Warren}},\ }\href {\doibase 10.1021/ic50107a012} {\bibfield  {journal}
  {\bibinfo  {journal} {Inorg. Chem}\ }\textbf {\bibinfo {volume} {11}},\
  \bibinfo {pages} {51} (\bibinfo {year} {1972})}\BibitemShut {NoStop}%
\bibitem [{\citenamefont {Magnuson}(1984)}]{Magnuson84}%
  \BibitemOpen
  \bibfield  {author} {\bibinfo {author} {\bibfnamefont {R.~H.}\ \bibnamefont
  {Magnuson}},\ }\href {\doibase 10.1021/ic00172a002} {\bibfield  {journal}
  {\bibinfo  {journal} {Inorg. Chem.}\ }\textbf {\bibinfo {volume} {23}},\
  \bibinfo {pages} {387} (\bibinfo {year} {1984})}\BibitemShut {NoStop}%
\bibitem [{\citenamefont {Reynolds}\ \emph {et~al.}(1992)\citenamefont
  {Reynolds}, \citenamefont {Delfs}, \citenamefont {Figgis}, \citenamefont
  {Henderson}, \citenamefont {Moubaraki},\ and\ \citenamefont
  {Murray}}]{Reynolds92}%
  \BibitemOpen
  \bibfield  {author} {\bibinfo {author} {\bibfnamefont {P.~A.}\ \bibnamefont
  {Reynolds}}, \bibinfo {author} {\bibfnamefont {C.~D.}\ \bibnamefont {Delfs}},
  \bibinfo {author} {\bibfnamefont {B.~N.}\ \bibnamefont {Figgis}}, \bibinfo
  {author} {\bibfnamefont {M.~J.}\ \bibnamefont {Henderson}}, \bibinfo {author}
  {\bibfnamefont {B.}~\bibnamefont {Moubaraki}}, \ and\ \bibinfo {author}
  {\bibfnamefont {K.~S.}\ \bibnamefont {Murray}},\ }\href {\doibase
  10.1039/dt9920002309} {\bibfield  {journal} {\bibinfo  {journal} {J. Chem.
  Soc., Dalton Trans.}\ ,\ \bibinfo {pages} {2309}} (\bibinfo {year}
  {1992})}\BibitemShut {NoStop}%
\bibitem [{\citenamefont {Bogdanov}\ \emph {et~al.}(2015)\citenamefont
  {Bogdanov}, \citenamefont {Katukuri}, \citenamefont {Romhanyi}, \citenamefont
  {Yushankhai}, \citenamefont {Kataev}, \citenamefont {Buechner}, \citenamefont
  {van~den Brink},\ and\ \citenamefont {Hozoi}}]{Bogdanov15}%
  \BibitemOpen
  \bibfield  {author} {\bibinfo {author} {\bibfnamefont {N.~A.}\ \bibnamefont
  {Bogdanov}}, \bibinfo {author} {\bibfnamefont {V.~M.}\ \bibnamefont
  {Katukuri}}, \bibinfo {author} {\bibfnamefont {J.}~\bibnamefont {Romhanyi}},
  \bibinfo {author} {\bibfnamefont {V.}~\bibnamefont {Yushankhai}}, \bibinfo
  {author} {\bibfnamefont {V.}~\bibnamefont {Kataev}}, \bibinfo {author}
  {\bibfnamefont {B.}~\bibnamefont {Buechner}}, \bibinfo {author}
  {\bibfnamefont {J.}~\bibnamefont {van~den Brink}}, \ and\ \bibinfo {author}
  {\bibfnamefont {L.}~\bibnamefont {Hozoi}},\ }\href {\doibase
  10.1038/ncomms8306} {\bibfield  {journal} {\bibinfo  {journal} {Nat.
  Commun.}\ }\textbf {\bibinfo {volume} {6}},\ \bibinfo {pages} {7306}
  (\bibinfo {year} {2015})}\BibitemShut {NoStop}%
\bibitem [{\citenamefont {Griffiths}\ and\ \citenamefont
  {Owen}(1954)}]{Griffiths54}%
  \BibitemOpen
  \bibfield  {author} {\bibinfo {author} {\bibfnamefont {J.~H.~E.}\
  \bibnamefont {Griffiths}}\ and\ \bibinfo {author} {\bibfnamefont
  {J.}~\bibnamefont {Owen}},\ }\href@noop {} {\bibfield  {journal} {\bibinfo
  {journal} {Proc. Royal Soc. A}\ }\textbf {\bibinfo {volume} {226}},\ \bibinfo
  {pages} {96} (\bibinfo {year} {1954})}\BibitemShut {NoStop}%
\bibitem [{\citenamefont {Schirmer}\ \emph {et~al.}(1984)\citenamefont
  {Schirmer}, \citenamefont {Forster}, \citenamefont {Hesse}, \citenamefont
  {Wohlecke},\ and\ \citenamefont {Kapphan}}]{Schirmer84}%
  \BibitemOpen
  \bibfield  {author} {\bibinfo {author} {\bibfnamefont {O.~F.}\ \bibnamefont
  {Schirmer}}, \bibinfo {author} {\bibfnamefont {A.}~\bibnamefont {Forster}},
  \bibinfo {author} {\bibfnamefont {H.}~\bibnamefont {Hesse}}, \bibinfo
  {author} {\bibfnamefont {M.}~\bibnamefont {Wohlecke}}, \ and\ \bibinfo
  {author} {\bibfnamefont {S.}~\bibnamefont {Kapphan}},\ }\href {\doibase
  10.1088/0022-3719/17/7/024} {\bibfield  {journal} {\bibinfo  {journal} {J.
  Phys. C: Solid State Phys.}\ }\textbf {\bibinfo {volume} {17}},\ \bibinfo
  {pages} {1321} (\bibinfo {year} {1984})}\BibitemShut {NoStop}%
\bibitem [{\citenamefont {Ganduglia-Pirovano}\ \emph
  {et~al.}(2007)\citenamefont {Ganduglia-Pirovano}, \citenamefont {Hofmann},\
  and\ \citenamefont {Sauer}}]{Ganduglia07}%
  \BibitemOpen
  \bibfield  {author} {\bibinfo {author} {\bibfnamefont {M.~V.}\ \bibnamefont
  {Ganduglia-Pirovano}}, \bibinfo {author} {\bibfnamefont {A.}~\bibnamefont
  {Hofmann}}, \ and\ \bibinfo {author} {\bibfnamefont {J.}~\bibnamefont
  {Sauer}},\ }\href {\doibase 10.1016/j.surfrep.2007.03.002} {\bibfield
  {journal} {\bibinfo  {journal} {‎Surf. Sci. Rep.}\ }\textbf {\bibinfo
  {volume} {62}},\ \bibinfo {pages} {219} (\bibinfo {year} {2007})}\BibitemShut
  {NoStop}%
\bibitem [{\citenamefont {Iakovleva}\ \emph {et~al.}(2015)\citenamefont
  {Iakovleva}, \citenamefont {Vavilova}, \citenamefont {Grafe}, \citenamefont
  {Zimmermann}, \citenamefont {Alfonsov}, \citenamefont {Luetkens},
  \citenamefont {Klauss}, \citenamefont {Maljuk}, \citenamefont {Wurmehl},
  \citenamefont {B\"uchner},\ and\ \citenamefont {Kataev}}]{Iakovleva15}%
  \BibitemOpen
  \bibfield  {author} {\bibinfo {author} {\bibfnamefont {M.}~\bibnamefont
  {Iakovleva}}, \bibinfo {author} {\bibfnamefont {E.}~\bibnamefont {Vavilova}},
  \bibinfo {author} {\bibfnamefont {H.-J.}\ \bibnamefont {Grafe}}, \bibinfo
  {author} {\bibfnamefont {S.}~\bibnamefont {Zimmermann}}, \bibinfo {author}
  {\bibfnamefont {A.}~\bibnamefont {Alfonsov}}, \bibinfo {author}
  {\bibfnamefont {H.}~\bibnamefont {Luetkens}}, \bibinfo {author}
  {\bibfnamefont {H.-H.}\ \bibnamefont {Klauss}}, \bibinfo {author}
  {\bibfnamefont {A.}~\bibnamefont {Maljuk}}, \bibinfo {author} {\bibfnamefont
  {S.}~\bibnamefont {Wurmehl}}, \bibinfo {author} {\bibfnamefont
  {B.}~\bibnamefont {B\"uchner}}, \ and\ \bibinfo {author} {\bibfnamefont
  {V.}~\bibnamefont {Kataev}},\ }\href {\doibase 10.1103/PhysRevB.91.144419}
  {\bibfield  {journal} {\bibinfo  {journal} {Phys. Rev. B}\ }\textbf {\bibinfo
  {volume} {91}},\ \bibinfo {pages} {144419} (\bibinfo {year}
  {2015})}\BibitemShut {NoStop}%
\bibitem [{\citenamefont {Ou}\ \emph {et~al.}(2014)\citenamefont {Ou},
  \citenamefont {Li}, \citenamefont {Fan}, \citenamefont {Wang},\ and\
  \citenamefont {Wu}}]{Ou14}%
  \BibitemOpen
  \bibfield  {author} {\bibinfo {author} {\bibfnamefont {X.}~\bibnamefont
  {Ou}}, \bibinfo {author} {\bibfnamefont {Z.}~\bibnamefont {Li}}, \bibinfo
  {author} {\bibfnamefont {F.}~\bibnamefont {Fan}}, \bibinfo {author}
  {\bibfnamefont {H.}~\bibnamefont {Wang}}, \ and\ \bibinfo {author}
  {\bibfnamefont {H.}~\bibnamefont {Wu}},\ }\href {\doibase
  {10.1038/srep07542}} {\bibfield  {journal} {\bibinfo  {journal} {Sci. Rep.}\
  }\textbf {\bibinfo {volume} {4}},\ \bibinfo {pages} {{7542}} (\bibinfo {year}
  {2014})}\BibitemShut {NoStop}%
\bibitem [{\citenamefont {Kanungo}\ \emph {et~al.}(2016)\citenamefont
  {Kanungo}, \citenamefont {Yan}, \citenamefont {Felser},\ and\ \citenamefont
  {Jansen}}]{Kanungo16}%
  \BibitemOpen
  \bibfield  {author} {\bibinfo {author} {\bibfnamefont {S.}~\bibnamefont
  {Kanungo}}, \bibinfo {author} {\bibfnamefont {B.}~\bibnamefont {Yan}},
  \bibinfo {author} {\bibfnamefont {C.}~\bibnamefont {Felser}}, \ and\ \bibinfo
  {author} {\bibfnamefont {M.}~\bibnamefont {Jansen}},\ }\href {\doibase
  10.1103/PhysRevB.93.161116} {\bibfield  {journal} {\bibinfo  {journal} {Phys.
  Rev. B}\ }\textbf {\bibinfo {volume} {93}},\ \bibinfo {pages} {161116}
  (\bibinfo {year} {2016})}\BibitemShut {NoStop}%
\bibitem [{Sup()}]{Suppl}%
  \BibitemOpen
  \href@noop {} {}\bibinfo {note} {See the attached Supplemental Material for
  the statistical analysis of the distribution of magnetic centers on a simple-
  and double-perovskite lattice.}\BibitemShut {Stop}%
\end{thebibliography}%

\onecolumngrid

\pagenumbering{gobble}

\hbox{ }
\vspace*{-2cm}

\hspace*{-2.3cm}
\includegraphics[page=1]{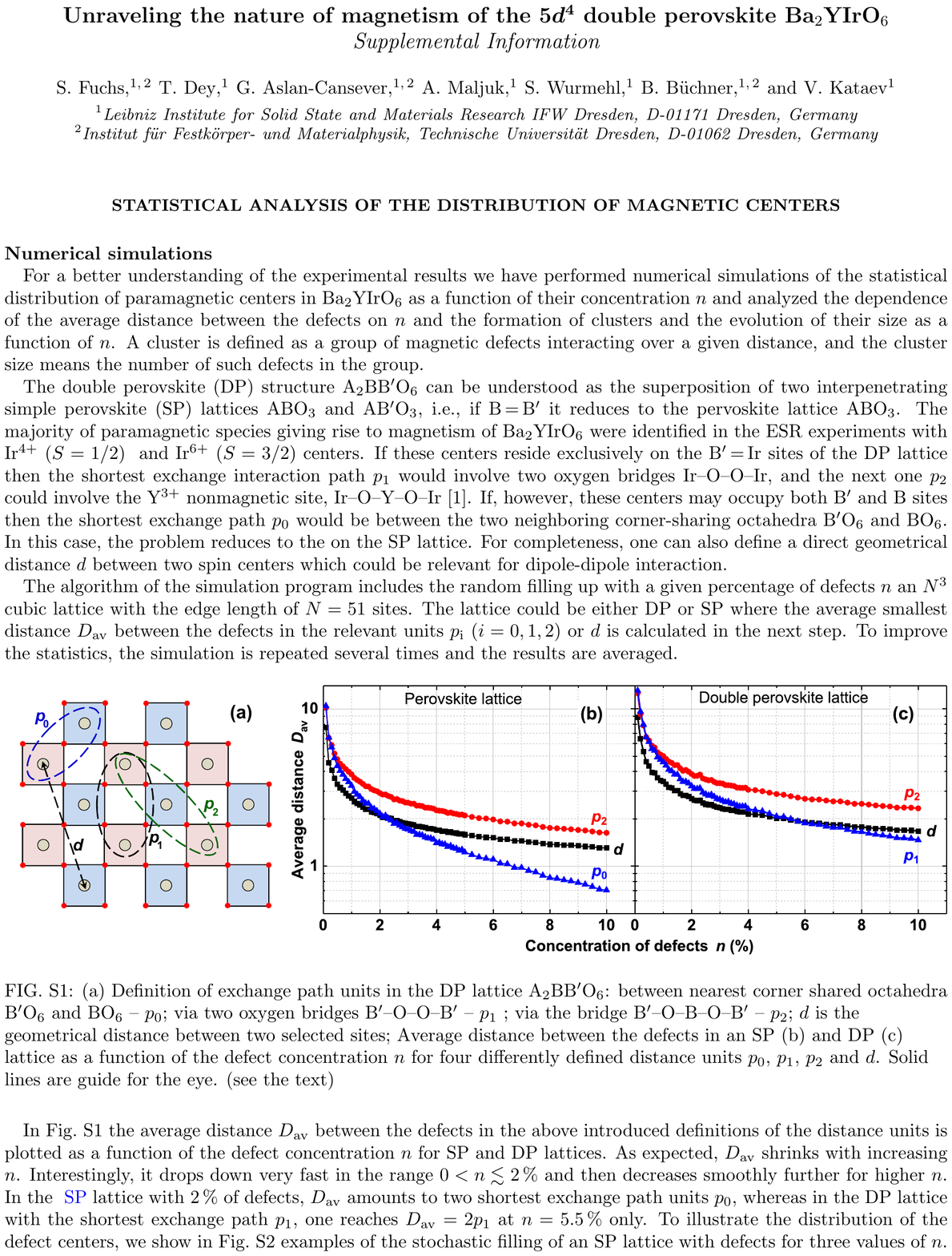}\\
\vspace*{-2cm}

\hspace*{-2.3cm}
\includegraphics[page=2]{Supplement_revision.pdf}\\
\vspace*{-2cm}

\hspace*{-2.3cm}
\includegraphics[page=3]{Supplement_revision.pdf}\\
\vspace*{-2cm}

\hspace*{-2.3cm}
\includegraphics[page=4]{Supplement_revision.pdf}\\
\vspace*{-2cm}

\end{document}